\documentclass[a4paper,11pt]{article}

\usepackage{jheppub}

\usepackage[T1]{fontenc}
\usepackage{graphicx}

\title{\boldmath  Mixed Inert Scalar Triplet Dark Matter, Radiative Neutrino Masses and Leptogenesis}

\author{Wen-Bin Lu}

\emailAdd{robinsonlu@sjtu.edu.cn}

\author{Pei-Hong Gu}

\emailAdd{peihong.gu@sjtu.edu.cn}

\affiliation{Department of Physics and Astronomy, Shanghai Jiao Tong University, \\800 Dongchuan Road, Shanghai 200240, China}

\abstract{The neutral component of an inert scalar multiplet with hypercharge can provide a stable dark matter particle when its real and imaginary parts have a splitting mass spectrum. Otherwise, a tree-level dark matter-nucleon scattering mediated by the $Z$ boson will be much above the experimental limit. In this paper we focus on a mixed inert scalar triplet dark matter scenario where a complex scalar triplet with hypercharge can mix with another real scalar triplet without hypercharge through their renormalizable coupling to the standard model Higgs doublet. We consider three specified cases that carry most of the relevant features of the full parameter space: (i) the neutral component of the real triplet dominates the dark matter particle, (ii) the neutral component of the complex triplet dominates the dark matter particle; and (iii) the neutral components of the real and complex triplets equally constitute the dark matter particle. Subject to the dark matter relic abundance and direct detection constraint, we perform a systematic study on the allowed parameter space with particular emphasis on the interplay among triplet-doublet terms and gauge interactions. In the presence of these mixed inert scalar triplets, some heavy Dirac fermions composed of inert fermion doublets can be utilized to generate a tiny Majorana neutrino mass term at one-loop level and realize a successful leptogenesis for explaining the cosmic baryon asymmetry.}

\begin{document}
\maketitle
\flushbottom

\section{Introduction}

Extensive astronomical and cosmological observations have offered overwhelming evidence for the existence of nonluminous dark matter (DM) \cite{olive2014}. In order to address its unknown composition and provenance, a profusion of well-motivated new physics have been proposed beyond the $SU(3)_c^{}\times SU(2)_L^{}\times U(1)_Y^{}$ standard model (SM) \cite{dmreview}. The particle nature in these new physics models also varies. The DM candidate can be a scalar, a fermion or a vector. In terms of a scalar particle to provide a viable DM, it first of all cannot develop a vacuum expectation value (VEV). Otherwise, the DM scalar will mix with the SM Higgs boson due to the Higgs portal interaction and eventually decay into the SM species. Furthermore, the DM scalar should also be forbidden to have any Yukawa couplings with the SM fermions. Therefore, it is much considered to exclude the unexpected VEV and/or Yukawa couplings by invoking a $Z_2^{}$ discrete symmetry, under which the DM scalar is odd while the SM fields are even. Such DM scalar is conventionally termed as an inert field.

Generically, a DM scalar can be either an SM singlet or a neutral component of an SM multiplet. For instance \cite{sz1985,mcdonald1994}, an inert scalar singlet can annihilate into the SM particles through a Higgs portal interaction and the correct thermal relic can be achieved. Alternatively, for an inert scalar multiplet, the electroweak radiative corrections tend to make the charged components slightly heavier than the neutral one, thus provide an automatically stable candidate to leave a relic density in the universe. The neutral scalar from a multiplet with non-zero hypercharge couples to the $Z$ boson so that it would have a much enhanced spin independent cross section above the present bounds. To evade this constraint, a viable multiplet DM entails either zero hypercharge, for instance the real triplet scenario \cite{cfs2006,fprw2009,agn2011}, or a mass splitting between the real and imaginary parts to efficiently suppress the inelastic scattering off nuclei. A well-known paradigm is the inert Higgs doublet model \cite{ma2006,bhr2006,hllr2009,inertdoublet} where the quartic coupling between the inert Higgs doublet and the SM Higgs doublet in the most general renormalizable potential inherently allows for such a mass splitting.

On the other hand, the atmospheric, solar, accelerator and reactor neutrino experiments have established the phenomena of neutrino oscillations \cite{olive2014}. This again calls for new physics to make three flavors of neutrinos massive and mixed. Meanwhile, the cosmological observations indicate that the neutrino masses should be below the eV scale \cite{olive2014}. Among others, the famous seesaw mechanism \cite{minkowski1977} is much acclaimed to naturally give the tiny neutrino masses. In various seesaw extensions of the SM \cite{minkowski1977,mw1980,flhj1989,ma1998,barr2004}, some heavy fields, which are responsible for highly suppressing the neutrino masses, can decay to generate a lepton asymmetry then converted to a baryon asymmetry by virtue of the sphaleron \cite{krs1985} processes. This is the so-called leptogenesis mechanism \cite{fy1986} and has been extensively studied \cite{lpy1986,fps1995,ms1998,bcst1999,hambye2001,di2002,gnrrs2003,hs2004,bbp2005,dnn2008,dhh2014,ksy2015,fmmn2015}. In the usual seesaw-leptogenesis context, the lepton number is explicitly broken and the neutrinos are of Majorana nature. For example, we have a lepton number violation of two units from the Majorana mass term of the fermion singlets(triplets) in the type-I(III) seesaw, or from the trilinear coupling of the Higgs triplets to the SM Higgs doublet in the type-II seesaw.

The particles for the DM may also play an essential role in the generation of the neutrino masses \cite{knt2003,ma2006,ma2006-2,kms2006,ms2009,chrt2011,ma2015,gms2016,mine,reftomine} and even the origin of the baryon asymmetry \cite{ma2006}. For example \cite{ma2006}, one extends the SM by two or more fermion singlets and a second Higgs doublet, which are all odd under an exactly conserved $Z_2^{}$ discrete symmetry, to simultaneously explain the puzzles of the neutrino masses, the baryon asymmetry and the DM. Specifically, the new Higgs doublet can provide a real scalar to be a stable DM particle, while the new fermions with heavy Majorana masses can highly suppress the radiative neutrino masses and their decays can realize a successful leptogenesis.

In this paper we shall explore a mixed inert scalar triplet DM scenario where a complex scalar triplet with non-zero hypercharge and a real scalar triplet with zero hypercharge mix with each other through their renormalizable coupling to the SM Higgs doublet. For demonstration we consider three limiting cases where (i) the neutral component of the real triplet dominates the dark matter particle, (ii) the neutral component of the complex triplet dominates the dark matter particle, and (iii) the neutral components of the real and complex triplets equally contribute to the dark matter particle. We will refer to these limiting cases as the dominant real triplet scenario, the dominant complex triplet scenario and the democratic real and complex triplet scenario. To accommodate the DM relic abundance and the DM direct detection constraint, we perform a systematic study on the allowed parameter space with particular emphasis on the interplay among triplet-doublet terms and gauge interactions. In what follows, we will introduce two types of fermion doublets with opposite hypercharges, which are also odd under the existent $Z_2^{}$ symmetry and are hence termed as the inert fermions, to construct some heavy Dirac fermions. Thanks to the Yukawa couplings of the inert scalar triplets and fermion doublets to the SM lepton doublets, we can realize a radiative generation of the Majorana neutrino masses and a successful leptogenesis for the cosmic baryon asymmetry. As we will show later, our leptogenesis possesses the new feature that the lepton asymmetry is induced by the decays of some heavy Dirac fermions rather than the usual Majorana ones.

In the following Section II, we will show the general picture of the mixed real and complex triplets. In Sections III, IV and V, we will study the DM particle arising from the dominant real triplet, the dominant complex triplet, as well as the democratic real and complex triplets, respectively.  In Section VI, we will demonstrate the generation of the neutrino masses and the baryon asymmetry. Finally Section VII is a summary. A few technical details are presented in the Appendices, including the explicit couplings of the individual components of the real and complex triplets to the SM species, the complete Feynman diagrams for the DM annihilations and co-annihilations, as well as the DM-nucleon scattering.

\section{Mixed inert scalar triplets}

Two inert scalars with different dimensions and/or hypercharges can mix with each other at the renormalizble level through their trilinear or quartic couplings with suitable Higgs scalars. By enumeration, we list the renormalizable, gauge-invariant terms for mixing the inert scalar singlet, doublet and triplet, in the presence of only one Higgs scalar, i.e. the SM Higgs doublet,
\begin{itemize}
\item singlet + doublet  (Model  I+II),
\begin{eqnarray}
\label{i+ii}
\mathcal{L}&\supset&-\rho\chi(\eta^\dagger_{}\phi+\textrm{H.c.})\,;
\end{eqnarray}
\item singlet + real triplet (Model I+IIIa),
\begin{eqnarray}
\label{i+iiia}
\mathcal{L}&\supset&-\lambda\chi(\phi^T_{}i\tau_2^{}\Sigma\tilde{\phi}+\textrm{H.c.})\,;
\end{eqnarray}
\item singlet + complex triplet (Model I+IIIb),
\begin{eqnarray}
\label{i+iiib}
\mathcal{L}&\supset&-\lambda\chi(\phi^T_{}i\tau_2^{}\Delta \phi+\textrm{H.c.})\,;
\end{eqnarray}
\item doublet + real triplet (Model II+IIIa),
\begin{eqnarray}
\label{ii+iiia}
\mathcal{L}&\supset&-\rho(\phi^T_{}i\tau_2^{}\Sigma\tilde{\eta}+\textrm{H.c.})\,;
\end{eqnarray}
\item doublet + complex triplet (Model II+IIIb),
\begin{eqnarray}
\label{ii+iiib}
\mathcal{L}&\supset&-\rho(\phi^T_{}i\tau_2^{}\Delta \eta+\textrm{H.c.})\,;
\end{eqnarray}
\item real triplet + complex triplet (Model IIIa+IIIb),
\begin{eqnarray}
\label{iiia+iiib}
\mathcal{L}&\supset&-\lambda(\phi^T_{}i\tau_2^{}\Sigma \Delta\phi+\textrm{H.c.})\,;
\end{eqnarray}
\end{itemize}
In the above we have denoted the SM Higgs doublet by
\begin{eqnarray}
\phi(1,2,-1/2)&=&\left[\begin{array}{c}\phi^0_{}\\
[2mm]
\phi^{-}_{}\end{array}\right]\,,
\end{eqnarray}
while the inert singlet, doublet and triplets by
\begin{eqnarray}
\chi(1,1,0)&=&\chi^\dagger_{}\,,\nonumber\\
[2mm]
\eta(1,2,-1/2)&=&\left[\begin{array}{c} \eta^0_{}\\
[2mm]
\eta^{-}_{}\end{array}\right]=\left[\begin{array}{c}\frac{1}{\sqrt{2}}(\eta^0_{R}+i\eta^0_{I})\\
[2mm]
\eta^{-}_{}\end{array}\right]\,,\nonumber\\
[2mm]
\Sigma(1,3,0)&=&\left[\begin{array}{cc}\frac{1}{\sqrt{2}}\sigma^0_{}&\sigma^+_{}\\
[2mm]
\sigma^-_{}&-\frac{1}{\sqrt{2}}\sigma^0_{}\end{array}\right]=\Sigma^\dagger_{}\,,\nonumber\\
[2mm]
\Delta(1,3,1)&=&\left[\begin{array}{cc}\frac{1}{\sqrt{2}}\delta^{+}_{}&\delta^{++}_{}\\
[2mm]
\delta^{0}_{}&-\frac{1}{\sqrt{2}}\delta^{+}_{}\end{array}\right]=\left[\begin{array}{cc}\frac{1}{\sqrt{2}}\delta^{+}_{}&\delta^{++}_{}\\
[2mm]
\frac{1}{\sqrt{2}}(\delta^{0}_{R}+i\delta^{0}_{I})&-\frac{1}{\sqrt{2}}\delta^{+}_{}\end{array}\right]\,.
\end{eqnarray}

Here and thereafter the brackets following the fields describe the transformations under the SM $SU(3)_c^{}\times SU(2)_L^{}\times U(1)_{Y}^{}$ gauge group. Note that more complicated models with complex scalars with vanishing hypercharge are equivalent to that with two interacting real scalars, simply doubling the number of degrees of freedom. Therefore without loss of generality, in the renormalizable models for mixed inert scalars shown in Eqs. (\ref{i+ii}-\ref{iiia+iiib}), we only take into account the cases of real singlets or multiplets.

While models involving a stable DM arising from two mixed inert scalars have not received the same level of attention as those with pure scalar singlet or multiplets, both the singlet-doublet and singlet-triplet combinations have been considered previously \cite{cheung,singlet-triplet}. Take the Model I+II in Eq. (\ref{i+ii}) as an example, the inert singlet $\chi$ and the real part $\eta^0_R$ of the neutral component $\eta^0_{}$ of the inert doublet $\eta$ mix after the electroweak symmetry breaking to produce two physical scalars, from which the lighter, also proved to be the lightest among other states, emerges as the DM candidate. The DM in the other models (\ref{i+iiia}-\ref{iiia+iiib}) for mixed inert scalars can be understood in a similar way. In the following of this paper, we will focus on the mixed inert scalar triplets, i.e. the model (\ref{iiia+iiib}).

\subsection{Real and complex inert triplets}

The Lagrangian involving the inert triplet scalars $\Sigma$ and $\Delta$ reads,
\begin{eqnarray}
\mathcal{L}&=&\frac{1}{2}\textrm{Tr}\left[ \left( D_\mu^{}\Sigma \right)^\dagger\left( D^\mu_{}\Sigma \right)\right] + \textrm{Tr}\left[ \left( D_\mu^{}\Delta \right)^\dagger\left( D^\mu_{}\Delta \right)\right] - V\,,\nonumber\\
&&
\end{eqnarray}
where the covariant derivatives are defined as
\begin{eqnarray}
D_\mu^{}\Sigma&=&\partial_\mu^{}\Sigma - i g \left[\frac{\tau_a^{}}{2}W_\mu^a, \Sigma \right]\,,\\
[2mm]
D_\mu^{}\Delta&=&\partial_\mu^{}\Delta - i g \left[\frac{\tau_a^{}}{2}W_\mu^a, \Delta \right] - g' B_\mu^{} \Delta\,,
\end{eqnarray}
with the corresponding general renormalizable scalar potential
\begin{eqnarray}
\label{potential}
V&=&\mu_\phi^2\phi^\dagger_{}\phi+\lambda_\phi^{}(\phi^\dagger_{}\phi)^2_{}+\frac{1}{2}M_\Sigma^2\textrm{Tr}(\Sigma^2_{})+\frac{1}{4}\lambda_\Sigma^{}[\textrm{Tr}(\Sigma^2_{}) ]^2_{}
+M_\Delta^2\textrm{Tr}(\Delta^\dagger_{}\Delta)+\lambda_\Delta^{}[\textrm{Tr}(\Delta^\dagger_{}\Delta)]^2_{}\nonumber\\
[2mm]
&&+\lambda'^{}_\Delta\textrm{Tr}[(\Delta^\dagger_{}\Delta)^2_{}]+\frac{1}{2}\kappa_1^{}\phi^\dagger_{}\phi\textrm{Tr}(\Sigma^2_{}) +\kappa^{}_2 \phi^\dagger_{}\phi \textrm{Tr}\left( \Delta^\dagger_{}\Delta \right) +\kappa^{}_3\phi^\dagger _{}\Delta\Delta^\dagger_{}\phi \nonumber\\
[2mm]
&&+\frac{1}{2} \kappa_4^{}\textrm{Tr}(\Sigma^2_{})\textrm{Tr}(\Delta^\dagger_{}\Delta)+\lambda\left(\phi^{T}i\tau_2^{}\Sigma\Delta\phi +\textrm{H.c.}\right)\,.\nonumber\\
&&
\end{eqnarray}
Here we have dropped the terms $\textrm{Tr}[\Delta^2_{}(\Delta^\dagger_{})^2_{}]$, $\phi^\dagger_{}\Sigma^2\phi$ and $\phi^\dagger\Delta^\dagger_{}\Delta\phi$ from the above potential, thanks to the identities $\textrm{Tr}[\Delta^2_{}(\Delta^\dagger_{})^2_{}]=\frac{1}{2}\textrm{Tr}(\Delta^2_{})\textrm{Tr}[(\Delta^\dagger_{})^2_{}]$, $\phi^\dagger_{}\Sigma^2_{}\phi=\frac{1}{2}\phi^\dagger_{}\phi\textrm{Tr}(\Sigma^2_{})$, $\textrm{Tr}[(\Delta^\dagger_{}\Delta)^2_{}] +\frac{1}{2}\textrm{Tr}(\Delta^2_{}) \textrm{Tr} [(\Delta^\dagger_{})^2_{}] = [\textrm{Tr}(\Delta^\dagger_{}\Delta)]^2_{}$ and $\phi^\dagger_{}\Delta^\dagger_{}\Delta\phi+\phi^\dagger_{}\Delta\Delta^\dagger_{}\phi=\phi^\dagger_{}\phi\textrm{Tr}(\Delta^\dagger_{}\Delta)$. Without loss of generality, it suffices to consider
\begin{eqnarray}
\lambda> 0\,,
\end{eqnarray}
since by applying a phase rotation of either $\Sigma \rightarrow -\Sigma$ or $\Delta \rightarrow -\Delta$, we can flip its sign. So the sign is unphysical. Furthermore, vacuum stability and perturbativity requirements imply
\begin{eqnarray}
&&0<\lambda,~\lambda_{\phi}^{},~\lambda_{\Sigma}^{},~\lambda^{}_\Delta+\lambda'^{}_{\Delta},~\lambda^{}_\Delta+\frac{1}{2}\lambda'^{}_{\Delta} <4\pi\,,\nonumber\\
[2mm]
&&-2\sqrt{\lambda_\phi^{}\lambda^{}_\Sigma}<\kappa_1^{}<4\pi\,,\nonumber\\
[2mm]
&&-2\sqrt{\lambda_\phi^{}(\lambda^{}_\Delta+\lambda'^{}_\Delta)}<\kappa^{}_2,~\kappa^{}_2+\kappa^{}_3<4\pi\,,\nonumber\\
[2mm]
&&-2\sqrt{\lambda_\phi^{}(\lambda^{}_\Delta+\frac{1}{2}\lambda'^{}_\Delta)}<\kappa^{}_2+\frac{1}{2}\kappa^{}_3<4\pi\,.
\end{eqnarray}

\subsection{Mass eigenstates}

After the SM Higgs doublet $\phi$ develops its VEV to spontaneously break the electroweak symmetry, it writes
\begin{eqnarray}
\phi=\left[\begin{array}{c}\frac{1}{\sqrt{2}}(h+v)\\
[2mm]
0\end{array}\right]~~\textrm{with}~~v=246\,\textrm{GeV}\,.
\end{eqnarray}
Here $h$ is the Higgs boson and with its mass
\begin{eqnarray}
m_h^{}=\sqrt{2\lambda_\phi^{}} v\simeq 125\,\textrm{GeV}~~\textrm{for}~~\lambda_\phi^{}\simeq 0.13\,.
\end{eqnarray}
At this stage, we can easily read the mass of the doubly charged component $\delta^{\pm\pm}_{}$ from the complex triplet $\Delta$, i.e.
\begin{eqnarray}
m_{\chi^{\pm\pm}_{}}^2=M_{\Delta}^2+\frac{1}{2}\kappa_2^{}v^2_{}+\frac{1}{2}\kappa_3^{}v^2_{}\,,~~(\delta^{\pm\pm}\equiv \chi^{\pm\pm})\,.
\end{eqnarray}
As for the neutral component $\sigma^0_{}$ from the real triplet $\Sigma$ and the other neutral component  $\delta^0_{}\equiv \frac{1}{\sqrt{2}}(\delta^0_R+i \delta^0_I)$ from the complex triplet $\Delta$, they will mix via the quartic coupling, i.e. the $\lambda$-term. Specifically, the squared mass matrix in the basis of the neutral fields $(\sigma^0_{}\,,~ \delta^0_R\,,~\delta^0_I)$ has the form
\begin{align}
\mathcal{M}_0^2=\left[\begin{array}{ccc} M_\Sigma^2+\frac{1}{2}\kappa_1^{} v^2_{} & -\frac{1}{2}\lambda v^2_{} & 0 \\
[1mm]\\
-\frac{1}{2}\lambda v^2_{} & M_\Delta^2+\frac{1}{2}\kappa_2^{} v^2_{} & 0 \\
[1mm]\\
0 & 0 &M_\Delta^2+\frac{1}{2}\kappa_2^{} v^2_{} \end{array}\right]\,.
\end{align}
The mass eigenstates should be
\begin{eqnarray}
\chi_1^0=\sigma^0_{}\sin\theta_0^{}+\delta^0_{R}\cos\theta_0^{}\,,\quad
\chi_2^0=\sigma^0_{}\cos\theta_0^{}-\delta^0_{R}\sin\theta_0^{}\,,\quad
\chi_3^0=\delta^0_I\,,
\end{eqnarray}
with the masses,
\begin{eqnarray}
m_{\chi_1^0}^2&=&\frac{1}{2}\left[\left(M_\Delta^2+\frac{1}{2}\kappa_2^{}v^2_{}\right)+\left(M_\Sigma^2+\frac{1}{2}\kappa_1^{}v^2_{}\right)\right]\nonumber\\
[2mm]&&-\frac{1}{2}\left\{\left[\left(M_\Delta^2+\frac{1}{2}\kappa_2^{}v^2_{}\right)-\left(M_\Sigma^2+\frac{1}{2}\kappa_1^{}v^2_{}\right)\right]^2_{}+\lambda^2_{}v^4_{}\right\}^{\frac{1}{2}}_{}\,,\nonumber\\
[2mm]
m_{\chi_2^0}^2&=&\frac{1}{2}\left[\left(M_\Delta^2+\frac{1}{2}\kappa_2^{}v^2_{}\right)+\left(M_\Sigma^2+\frac{1}{2}\kappa_1^{}v^2_{}\right)\right]\nonumber\\
[2mm]&&+\frac{1}{2}\left\{\left[\left(M_\Delta^2+\frac{1}{2}\kappa_2^{}v^2_{}\right)-\left(M_\Sigma^2+\frac{1}{2}\kappa_1^{}v^2_{}\right)\right]^2_{}+\lambda^2_{}v^4_{}\right\}^{\frac{1}{2}}_{}\,,\nonumber\\
[2mm]
m_{\chi_3^0}^2&=&M_\Delta^2+\frac{1}{2}\kappa_2^{} v^2_{}\,,
\end{eqnarray}
as well as the mixing angle,
\begin{eqnarray}
\tan 2\theta_0^{}=-\frac{\lambda v^2_{}}{(M_\Delta^2+\frac{1}{2}\kappa_2^{}v^2_{})-(M_\Sigma^2+\frac{1}{2}\kappa_1^{}v^2_{})}\,.
\end{eqnarray}
Similarly, the two singly charged components $(\sigma^{\pm}_{}\,,~\delta^{\pm}_{})$ are also mixed via the matrix
\begin{align}
\mathcal{M}_{\pm}^2=\left[\begin{array}{ccc} M_\Sigma^2+\frac{1}{2}\kappa_1^{} v^2_{} & \frac{1}{2\sqrt{2}}\lambda v^2_{}
\\[1mm]
\\ \frac{1}{2\sqrt{2}}\lambda v^2_{} & M_\Delta^2+\frac{1}{2}\kappa_2^{}v^2_{}+\frac{1}{4}\kappa_3^{}v^2_{} \end{array}\right]\,,
\end{align}
from which we read two mass eigenstates,
\begin{eqnarray}
\chi_1^\pm=\sigma^\pm_{}\sin\theta_\pm^{}+\delta^\pm_{}\cos\theta_\pm^{}\,,\quad
\chi_2^\pm=\sigma^\pm_{}\cos\theta_\pm^{}-\delta^\pm_{}\sin\theta_\pm^{}\,,
\end{eqnarray}
with the masses,
\begin{eqnarray}
m_{\chi_1^\pm}^2&=&\frac{1}{2}\left[\left(M_\Delta^2+\frac{1}{2}\kappa_2^{}v^2_{}+\frac{1}{4}\kappa_3^{}v^2_{}\right)+\left(M_\Sigma^2+\frac{1}{2}\kappa_1^{}v^2_{}\right)\right]\nonumber\\
&&-\frac{1}{2}\left\{\left[\left(M_\Delta^2+\frac{1}{2}\kappa_2^{}v^2_{}+\frac{1}{4}\kappa_3^{}v^2_{}\right)-\left(M_\Sigma^2+\frac{1}{2}\kappa_1^{}v^2_{}\right)\right]^2_{}+\frac{1}{2}\lambda^2_{}v^4_{}\right\}^{\frac{1}{2}}_{}\,,\nonumber\\
[2mm]
m_{\chi_2^\pm}^2&=&\frac{1}{2}\left[\left(M_\Delta^2+\frac{1}{2}\kappa_2^{}v^2_{}+\frac{1}{4}\kappa_3^{}v^2_{}\right)+\left(M_\Sigma^2+\frac{1}{2}\kappa_1^{}v^2_{}\right)\right]\nonumber\\
&&+\frac{1}{2}\left\{\left[\left(M_\Delta^2+\frac{1}{2}\kappa_2^{}v^2_{}+\frac{1}{4}\kappa_3^{}v^2_{}\right)-\left(M_\Sigma^2+\frac{1}{2}\kappa_1^{}v^2_{}\right)\right]^2_{}+\frac{1}{2}\lambda^2_{}v^4_{}\right\}^{\frac{1}{2}}_{}\,,\end{eqnarray}
and the mixing angle,
\begin{eqnarray}
\tan 2\theta_\pm^{}=\frac{\frac{1}{\sqrt{2}}\lambda v^2_{}}{[M_\Delta^2+\frac{1}{2}(\kappa_2^{}+\frac{1}{2}\kappa_3^{})v^2_{}]-(M_\Sigma^2+\frac{1}{2}\kappa_1^{}v^2_{})}\,.
\end{eqnarray}

\begin{table*}[tbp]
\centering
\resizebox{0.95\hsize}{!}{
\begin{tabular}{|l|l|}
\hline
~&~\\
\quad\quad \quad Mass eigenstate   &\quad\quad \quad\quad\quad \quad \quad\quad\quad \quad\quad \quad Mass square \\
~&~\\
\hline\hline
$~$&$~$\\
$~~\tan 2\theta_0^{}=-\frac{\lambda v^2_{}}{m_\Delta^2-m_\Sigma^2}~~$&$~$\\
$~$&$~$\\
$~~\chi^0_1=\sigma^0_{}\sin\theta_0^{}+\delta^0_R\cos\theta_0^{} ~~$  &  $ ~~m_{\chi_1^0}^2=\frac{1}{2}\left(m_\Delta^2+m_\Sigma^2\right)
-\frac{1}{2}\sqrt{\left(m_\Delta^2-m_\Sigma^2\right)^2_{}+\lambda^2_{}v^4_{}}~~$ \\
$~$&$~$\\
$ ~~\chi^0_2=\sigma^0_{}\cos\theta_0^{}-\delta^0_{R}\sin\theta_0^{}~~$  &  $~~m_{\chi_2^0}^2=\frac{1}{2}\left(m_\Delta^2+m_\Sigma^2\right)
+\frac{1}{2}\sqrt{\left(m_\Delta^2-m_\Sigma^2\right)^2_{}+\lambda^2_{}v^4_{}}~~$ \\
$~$&$~$\\
$~~\chi^0_3=\delta^0_I~~$ &$~~ m_{\chi_3^0}^2=m_\Delta^2~~$\\
$~$&$~$\\
\hline
$~$&$~$\\
$~~\tan 2\theta_\pm^{}=\frac{\frac{1}{\sqrt{2}}\lambda v^2_{}}{(m_\Delta^2+\frac{1}{4}\kappa_3^{}v^2_{})-m_\Sigma^2}~~$&$~$\\
$~$&$~$\\
$~~ \chi_1^\pm= \sigma^\pm_{}\sin\theta_\pm^{}+\delta^\pm_{}\cos\theta_\pm^{}~~$ &  $~~m_{\chi^\pm_1}^2=\frac{1}{2}\left[\left(m_\Delta^2+\frac{1}{4}\kappa_3^{}v^2_{}\right)+m_\Sigma^2\right]-\frac{1}{2}\sqrt{\left[\left(m_\Delta^2+\frac{1}{4}\kappa_3^{}v^2_{}\right)-m_\Sigma^2\right]^2_{}+\frac{1}{2}\lambda^2_{}v^4_{}}~~$\\
$~$&$~$\\
$~~ \chi_2^\pm=\sigma^\pm_{}\cos\theta_\pm^{}-\delta^{\pm}_{}\sin\theta_\pm^{} ~~$ &  $~~m_{\chi^\pm_2}^2=\frac{1}{2}\left[\left(m_\Delta^2+\frac{1}{4}\kappa_3^{}v^2_{}\right)+m_\Sigma^2\right]+\frac{1}{2}\sqrt{\left[\left(m_\Delta^2+\frac{1}{4}\kappa_3^{}v^2_{}\right)-m_\Sigma^2\right]^2_{}+\frac{1}{2}\lambda^2_{}v^4_{}}~~$ \\
$~$&$~$\\
\hline
$~$&$~$\\
$~~\chi^{\pm\pm}_{}= \delta^{\pm\pm}_{}~~$  & $ ~~m_{\chi^{\pm\pm}_{}}^2=m_\Delta^2+\frac{1}{2}\kappa_3 v^2~~$ \\
$~$&$~$\\
\hline
\end{tabular}}  
\caption{\label{mass-table} Mass eigenstates from the mixed triplet scalars $\Sigma$ and $\Delta$.}
\end{table*}

We summarize the six physical states including three neutral $\chi^{0}_{1,2,3}$, two singly charged $\chi^{\pm}_{1,2}$ and a doubly charged $\chi^{\pm\pm}_{}$ in Table~\ref{mass-table}, where we have conveniently denoted
\begin{eqnarray}
m_\Sigma^2\equiv M_\Sigma^2+\frac{1}{2}\kappa_1^{}v^2_{}\,,~~m_\Delta^2\equiv M_\Delta^2+\frac{1}{2}\kappa_2^{}v^2_{}\,.
\end{eqnarray}
Apparently, one can invariantly arrive at one neutral that is lighter than all the other states, i.e.
\begin{eqnarray}
m_{\chi^0_{1}}^2<m_{\chi^0_{2,3}}^2\,,~m_{\chi^{\pm}_{1,2}}^2\,,~m_{\chi^{\pm\pm}_{}}^2\,,
\end{eqnarray}
by choosing appropriate quartic couplings $\kappa_{1,2,3}$ and $\lambda$. This lightest neutral field $\chi_1^0$ emerges as the DM candidate, while the other mass eigenstates will eventually decay into this DM particle with certain SM species, hence contributing to the DM relic density \cite{planck},
\begin{eqnarray}
\Omega_{\text{DM}}h^2=0.1188\pm0.0010\,,
\end{eqnarray}
through their co-annihilations with the DM particle. We will discuss explicitly later.

\subsection{Limiting cases}

Under some limiting conditions, mass eigenstates and their corresponding masses are greatly simplified. We choose to anatomize three limiting cases with $m_\Delta^2\gg m_\Sigma^2 \gg v^2_{}$, $m_\Sigma^2\gg m_\Delta^2 \gg v^2_{}$ and  $m_\Sigma^2= m_\Delta^2\equiv m^2_{} \gg v^2_{}$, that respectively read
\begin{itemize}
\item $\theta_0^{}\simeq -\frac{\pi}{2}\,,~\theta_{\pm}^{}\simeq \frac{\pi}{2}$ and
\begin{eqnarray}
\label{real}
\begin{array}{lcl}
~~\,\chi^0_1\simeq -\sigma^0_{} &~\textrm{with}~&~~m_{\chi^0_1}^2\simeq m_\Sigma^2 -\frac{\lambda^2_{}v^4_{}}{4m_\Delta^2}\,,\\
[5mm]
~~\,\chi^0_{2}\simeq \delta^0_{R}&~\textrm{with}~&~~m_{\chi^0_2}^2\simeq m_\Delta^2\,,\\
[5mm]
~~\,\chi^0_{3}= \delta^0_{I}&~\textrm{with}~&~~m_{\chi^0_3}^2 = m_\Delta^2\,,\\
[5mm]
~~\chi^{\pm}_{1}\simeq -\sigma^{\pm}_{}&~\textrm{with}~&~\,m_{\chi^\pm_1}^2\simeq m_\Sigma^2 -\frac{\lambda^2_{}v^4_{}}{8m_\Delta^2}\,,\\
[5mm]
~~\chi^{\pm}_{2}\simeq \delta^{\pm}_{}&~\textrm{with}~&~\,m_{\chi^\pm_2}^2\simeq m_\Delta^2\,,\\
[5mm]
\chi^{\pm\pm}_{}= \sigma^{\pm\pm}_{}&~\textrm{with}~&m_{\chi^{\pm\pm}_{}}^2\simeq m_\Delta^2\,;\\
\end{array}
\end{eqnarray}
\item $\theta_0^{}\simeq 0\,,~\theta_{\pm}^{}\simeq \pi$ and
\begin{eqnarray}
\label{complex}
\begin{array}{lcl}
~~\,\chi^0_1\simeq \delta^0_{R} &~\textrm{with}~&~~m_{\chi^0_1}^2\simeq m_\Delta^2 -\frac{\lambda^2_{}v^4_{}}{4m_\Sigma^2}\,,\\
[5mm]
~~\,\chi^0_{2}\simeq \sigma^0_{}&~\textrm{with}~&~~m_{\chi^0_2}^2\simeq m_\Sigma^2\,,\\
[5mm]
~~\,\chi^0_{3}= \delta^0_{I}&~\textrm{with}~&~~m_{\chi^0_3}^2 = m_\Delta^2\,,\\
[5mm]
~~\chi^{\pm}_{1}\simeq -\delta^{\pm}_{}&~\textrm{with}~&~\,m_{\chi^\pm_1}^2\simeq m_\Delta^2 +\frac{1}{4}\kappa_3^{}v^2_{}-\frac{\lambda^2_{}v^4_{}}{8m_\Sigma^2}\,,\\
[5mm]
~~\chi^{\pm}_{2}\simeq -\sigma^{\pm}_{}&~\textrm{with}~&~\,m_{\chi^\pm_2}^2\simeq m_\Sigma^2\,,\\
[5mm]
\chi^{\pm\pm}_{}= \sigma^{\pm\pm}_{}&~\textrm{with}~&m_{\chi^{\pm\pm}_{}}^2= m_\Delta^2+\frac{1}{2}\kappa_3^{}v^2_{}\,;\\
\end{array}
\end{eqnarray}
\item $\theta_0^{}=\frac{\pi}{4}\,,~\theta_{\pm}^{}=\frac{1}{2}\arctan\left(\frac{\lambda}{4\sqrt{2}\kappa_3^{}}\right)$ and
\begin{eqnarray}
\label{realcomplex}
\begin{array}{lcl}
~~\,\chi^0_1=\frac{1}{\sqrt{2}}\left( \sigma^0_{}+\delta^0_{R} \right)&\textrm{with}&~~m_{\chi^0_1}^2= m_{}^2 -\frac{1}{2}\lambda v^2_{}\,,\\
[5mm]
~~\,\chi^0_{2}=\frac{1}{\sqrt{2}}\left( \sigma^0_{}-\delta^0_{R} \right) &\textrm{with} & ~~m_{\chi^0_2}^2= m_{}^2 +\frac{1}{2}\lambda v^2_{}\,,\\
[5mm]
~~\,\chi^0_{3}= \delta^0_{I}& \textrm{with}& ~~m_{\chi^0_3}^2 = m_{}^2\,,\\
[5mm]
~~\chi^{\pm}_{1}= \sigma^\pm_{}\sin\theta_\pm^{}+\delta^\pm_{}\cos\theta_\pm^{}&\textrm{with}&~\,m_{\chi^\pm_1}^2= m_{}^2 +\frac{1}{2}\left(\frac{1}{4}\kappa_3^{}-\sqrt{\frac{1}{16}\kappa_3^2+\frac{1}{2}\lambda^2_{}}\right)v^2_{}\,,\\
[5mm]
~~\chi^{\pm}_{2}=\sigma^\pm_{}\cos\theta_\pm^{}-\delta^{\pm}_{}\sin\theta_\pm^{} &\textrm{with}&~\,m_{\chi^\pm_2}^2= m_{}^2 +\frac{1}{2}\left(\frac{1}{4}\kappa_3^{}+\sqrt{\frac{1}{16}\kappa_3^2+\frac{1}{2}\lambda^2_{}}\right)v^2_{}\,,\\
[5mm]
\chi^{\pm\pm}_{}= \sigma^{\pm\pm}_{}&\textrm{with}&m_{\chi^{\pm\pm}_{}}^2= m_\Delta^2+\frac{1}{2}\kappa_3^{}v^2_{}\,.
\end{array}
\end{eqnarray}
\end{itemize}

As shown in Eqs. (\ref{real}-\ref{realcomplex}), for certain parameter choice, the neutral $\sigma^0_{}$ from the real triplet $\Sigma$ and the another neutral $\delta^0_R$ from the complex triplet $\Delta$ can give a dominant or an equal contribution to the lightest physical state $\chi^0_1$, i.e. the DM particle. We will refer to these three limiting cases as the dominant real triplet, the dominant complex triplet and the democratic real and complex triplets, respectively.

\section{Dominant real triplet}

Admittedly, a pure inert real triplet scalar can provide a viable DM candidate, as has been incorporated in the so-called minimal DM framework \cite{cfs2006}. In the dominant real triplet scenario (\ref{real}), the DM scalar $\chi^0_1$ is comprised mostly of $\sigma^0_{}$, the neutral component of the real triplet $\Sigma$, and the lightest charged scalar $\chi^{\pm}_{1}$ mostly of the charged component $\sigma^{\pm}_{}$. Moreover, the mass splitting between $\chi^0_1$ and $\chi^{\pm}_{1}$ is much smaller than their masses themselves. Actually, we can easily read
\begin{eqnarray}
\label{rmsplit}
\Delta m_\Sigma^{\pm}&=& m_{\chi^{\pm}_{1}}^{}-m_{\chi^{0}_{1}}^{}\simeq\frac{\lambda^2_{}v^4_{}}{16m_\Sigma^{} m_\Delta^2}+\frac{g^2_{}}{4\pi}m_W^{}\sin^2_{}\frac{\theta_W^{}}{2}\nonumber\\
[2mm]
&\simeq & 45\,\textrm{MeV} \left(\frac{\lambda}{4\pi}\right)^2_{}\left(\frac{10\,m_{\Sigma}^{}}{m_\Delta^{}}\right)^2_{}\left(\frac{2\,\textrm{TeV}}{m_\Sigma^{}}\right)+166\,\textrm{MeV}\,,
\end{eqnarray}
with the second part being the electroweak loop correction. It should be noted that although allowed to be rather small, the mixing strength $\lambda$ must be nonzero since otherwise, the neutral component $\delta^0_{}$ of the complex triplet $\Delta$ would also independently keep stable and contribute to the DM relic abundance. While unfortunately, this DM scenario has been excluded by the DM direct detection experiments.

\subsection{Relic density}

For the highly quasi-degenerate states $\chi^0_1\simeq \sigma^0_{}$ and $\chi^{\pm}_{1}\simeq -\sigma^{\pm}_{}$, all the following annihilation and co-annihilation channels should play a significant role in determining the DM relic density \cite{boltzmann,gg1991},
\begin{eqnarray}
\label{acoreal}
\sigma^{0}_{}\sigma^{0}_{}&\rightarrow& W^{+}_{}W^{-}_{}\,,~\phi^{\ast}_{}\phi^{}_{}\,;\nonumber\\
[2mm]
\sigma^{0}_{}\sigma^{\pm}_{}&\rightarrow& W^{3}_{}W^{\pm}_{}\,,~f'\bar{f}\,,~\phi^{\ast}_{}\phi^{}_{}\,;\nonumber\\
[2mm]
\sigma^{+}_{}\sigma^{-}_{}&\rightarrow& W^{+}_{}W^{-}_{}\,,~f\bar{f}\,,~\phi^{\ast}_{}\phi^{}_{}\,;\nonumber\\
[2mm]
\sigma^{\pm}_{}\sigma^{\pm}_{}&\rightarrow& W^{\pm}_{}W^{\pm}_{}\,,
\end{eqnarray}
where $f$ and $f'$ denote the SM fermions. The Feynman diagrams for all the relevant processes are depicted in the figures of appendix \ref{ann-feyndiagram}

Up to the $p$-wave contributions, the thermally averaged cross sections are computed for each channel,
\begin{eqnarray}
\langle\sigma_{\sigma^{0}_{}\sigma^{0}_{}}^{} v_{\textrm{rel}}^{}\rangle&=&\frac{g_{}^4}{4\pi m_\Sigma^2}\left(1-\frac{5}{x}\right)+\frac{\kappa^2_{1}}{16\pi m_\Sigma^2}\left(1-\frac{3}{x}\right)\,, \nonumber \\
[2mm]
\langle\sigma_{\sigma^{0}_{}\sigma^{\pm}_{}}^{} v_{\textrm{rel}}^{}\rangle&=&\frac{g_{}^4}{16\pi m_\Sigma^2}\left(1-\frac{5}{4x}\right) \,,\nonumber \\
[2mm]
\langle\sigma_{\sigma^{+}_{}\sigma^{-}_{}}^{} v_{\textrm{rel}}^{}\rangle&=&\frac{3g_{}^4}{16\pi m_\Sigma^2}\left(1-\frac{15}{4x}\right)+\frac{\kappa^2_{1}}{16\pi m_\Sigma^2} \left(1-\frac{3}{x}\right)\,, \nonumber \\
[2mm]
\langle\sigma_{\sigma^{\pm}_{}\sigma^{\pm}_{}}^{} v_{\textrm{rel}}^{}\rangle&=&\frac{g_{}^4}{8\pi m_\Sigma^2}\left(1-\frac{5}{x}\right)\,,
\end{eqnarray}
and the effective cross section is given by \cite{st2003},
\begin{eqnarray}
\label{effsecr}
\langle\sigma_{\textrm{eff}}^{\Sigma} v_{\textrm{rel}}^{}\rangle&=&\frac{1}{9}\langle\sigma_{\sigma^{0}_{}\sigma^{0}_{}}^{} v_{\textrm{rel}}^{}\rangle+
\frac{4}{9}\langle\sigma_{\sigma^{0}_{}\sigma^{\pm}_{}}^{} v_{\textrm{rel}}^{}\rangle+
\frac{2}{9}\langle\sigma_{\sigma^{+}_{}\sigma^{-}_{}}^{} v_{\textrm{rel}}^{}\rangle+
\frac{2}{9}\langle\sigma_{\sigma^{\pm}_{}\sigma^{\pm}_{}}^{} v_{\textrm{rel}}^{}\rangle\nonumber\\
[2mm]
&=&\frac{g_{}^4}{8\pi m_\Sigma^2}\left(1-\frac{15}{4 x}\right)+\frac{\kappa^2_{1}}{48\pi m_\Sigma^2}\left(1-\frac{3}{x}\right)\,.
\end{eqnarray}
where we have defined
\begin{eqnarray}
x\equiv \frac{m_\Sigma^{}}{T}\,.
\end{eqnarray}
Technically, the DM number density $n$ could be traced by solving the exact Boltzmann equation,
\begin{eqnarray}
\label{boltzmanneqn}
\frac{dn}{dt}+3Hn\,=\,-\langle\sigma_{\textrm{eff}}^{\Sigma}v_{\textrm{rel}}^{}\rangle\left(n^2-n_{\textrm{eq}}^2\right)\,,
\end{eqnarray}
with $H$ being the Hubble constant while $n_{\textrm{eq}}^{}$ being the equilibrium number density. From the above Boltzmann equation, a reliable approximation is obtained to yield the final DM relic abundance \cite{boltzmann,kt1990},
\begin{eqnarray}\label{relicdensity}
\Omega_{\textrm{DM}}^{}h^2_{}\simeq\frac{1.07\times10^9_{}\textrm{GeV}^{-1}_{}}{J(x_F^{}) g_{\ast}^{1/2} M_{\textrm{Pl}}^{}}\,.
\end{eqnarray}
Here and thereafter $M_\textrm{Pl}^{}\simeq 1.22\times 10^{19}\,\textrm{GeV}$ is the Planck mass, $g_{\ast}^{}\simeq 106.75$ is the number of the relativistic degrees of freedom at the freeze-out point, while $J(x_F^{})$ is an integral,
\begin{eqnarray}
J(x_F^{})&=&\int_{x_F^{}}^\infty \frac{\langle\sigma_{\textrm{eff}}^{\Sigma}v_{\textrm{rel}}^{}\rangle}{x^2_{}} dx\,,
\end{eqnarray}
determined by the freeze-out point,
\begin{eqnarray}
\label{freezeout}
x_F^{}=\ln \frac{3\times 0.038\,M_{\textrm{Pl}}^{}m_\Sigma^{}\langle\sigma_{\textrm{eff}}^{\Sigma}v_{\textrm{rel}}^{}\rangle}{g_\ast^{1/2} x_F^{1/2}}\,,
\end{eqnarray}
at which the annihilations and co-annihilations become slower than the expansion rate of the universe.

\begin{figure}[htp]
  \centering
  \includegraphics[width=8cm]{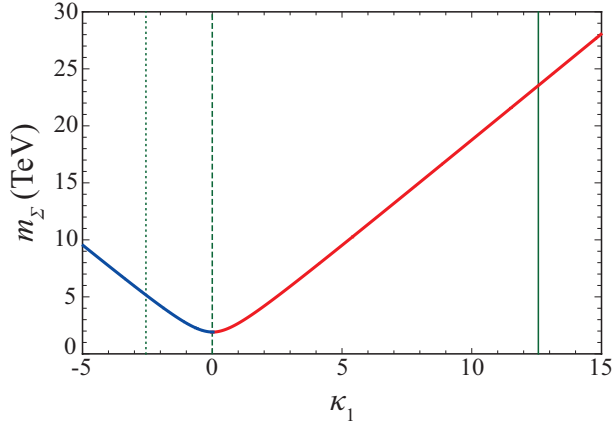}
  \caption{The correlation between the dark matter mass $m_\Sigma^{}$ and the Higgs portal coupling $\kappa_1^{}$. The dotted, dashed and solid vertical lines correspond to $\kappa_1^{}=-2.6$, $0$ and $4\pi$, respectively. The DM mass $m_\Sigma^{}$ will decrease from $m_\Sigma^{}=5.2\,\textrm{TeV}$ to $m_\Sigma^{}=2\,\textrm{TeV}$ when $\kappa_1^{}$ increases from $\kappa_1^{}=-2.6$ to $\kappa_1^{}=0$, subsequently, $m_\Sigma^{}$ will increase to $m_\Sigma^{}=23.6\,\textrm{TeV}$ when $\kappa_1^{}$ increases to $\kappa_1^{}=4\pi$.}
  \label{SigmaPureRelic}
\end{figure}

We remark on some quite typical features regarding the annihilations and co-annihilations in a scalar DM scenario. The first situation, which is commonly mentioned in literature, is that in the case of scalar DM, gauge-induced annihilations into the SM fermions and scalars are always $p$-wave suppressed, i.e. their thermally averaged cross sections have no $s$-wave contributions. Although the large number of fermions present in the SM could partially compensate this suppression, this proves to be weakly manifest in the final result.

Another situation is the quantitative significance to include co-annihilations rather than the annihilations alone. As noted in \cite{boltzmann}, just having extra particles near in mass to a DM candidate will not make a big difference so long as the cross sections are similar. So even though we should take co-annihilations into serious account, the magnitude of effects could vary, either additive or suppressive is possible. This is because the total effective cross section should be nothing but a weighted average of the annihilation and co-annihilation cross sections, as in Eq. (\ref{effsecr}). Consequently, more co-annihilating channels will not always lead to a bigger effective cross section hence a smaller relic density, as naively expected.

These features are well encountered in our case here. Co-annihilation effect actually increases the relic density roughly by a factor of 2 in the gauge interactions and 3 in the quartic interactions compared to the pure annihilation result.

Understandably, although the mass splitting in Eq. (\ref{rmsplit}) is different from that in the minimal DM scenario, the present real triplet scenario and the minimal DM scenario will give the same prediction on the correlation between the DM mass $m_\Sigma^{}$ and the Higgs portal coupling $\kappa_1^{}$. The relevant result is shown in Fig.~\ref{SigmaPureRelic}. Specifically, $m_\Sigma^{}$ will decrease from $m_\Sigma^{}=5.2\,\textrm{TeV}$ to $m_\Sigma^{}=2\,\textrm{TeV}$ when $\kappa_1^{}$ increases from $\kappa_1^{}=-2.6$ to $\kappa_1^{}=0$, subsequently, $m_\Sigma^{}$ will increase to $m_\Sigma^{}=23.6\,\textrm{TeV}$ when $\kappa_1^{}$ increases to $\kappa_1^{}=4\pi$.

\subsection{Direct detection}

An elastic cross section for the DM scalar $\sigma^0_{}$ to scatter off nuclei is generated at both tree and loop level. And the only tree level interaction proceeds through the exchange of the Higgs boson. For a display of Feynman diagrams for elastic scattering, see appendix \ref{ann-feyndiagram}. The DM-nucleon scattering is induced by the effective Lagrangian,
\begin{eqnarray}
\mathcal{L}&\supset&\frac{g^4_{}}{16\pi}\frac{m_{\Sigma}^{}}{m_W^{}}\left[\frac{1}{m_W^2}+\frac{1}{m_h^2}\left(1-\frac{16\pi }{g^4_{}} \frac{m_W^{}}{m_\Sigma^{}}\kappa_1^{}\right)\right] \sigma^0_{}\sigma^0_{}\sum_{q}^{}m_q^{}\bar{q}q~~\textrm{for}~~m_\Sigma^{} \gg m_W^{} \gg m_q^{}\,.\nonumber\\
&&
\end{eqnarray}
which contains only two unknown parameters: the DM mass $m_\Sigma^{}$ and the Higgs portal coupling $\kappa_1^{}$. Actually, we have shown in Fig. \ref{SigmaPureRelic} that the parameters $m_\Sigma^{}$ and
$\kappa_1^{}$ should be correlated by the DM relic density. The spin-independent DM-nucleon scattering cross section is given by
\begin{eqnarray}
\label{si}
\sigma_{\textrm{SI}}^{}=\frac{g^8_{}}{256\pi^3_{}} \frac{f_N^2 m_N^4}{m_W^2} \left[\frac{1}{m_W^2}+\frac{1}{m_h^2} \left( 1-\frac{16\pi}{g^4_{}}\frac{m_W^{}}{m_\Sigma^{}}\kappa_1^{}\right)
\right]^2~~\textrm{for}~~m_\Sigma^{} \gg m_W^{} \gg m_N^{}\,.
\end{eqnarray}
Here $m_N^{}\simeq 1\,\textrm{GeV}$ is the nucleon mass, while $f_N^{}\simeq 0.3$ is the effective coupling of the Higgs boson to the nucleon \cite{nucl_factor}, i.e.
\begin{eqnarray}\label{formfactor}
f_N^{}&=&\frac{1}{m_N^{}}\langle N|\sum_{q}^{}m_q^{}\bar{q}q|N\rangle\nonumber\\
[2mm]&=&\sum_{q=u,d,s}^{}f^{(N)}_{Tq}+\frac{2}{27}\sum_{q=c,b,t}^{}f^{(N)}_{Tq}~~\textrm{with}\nonumber\\
[2mm]
&&f_{Tu}^{(p)}=0.020\,,~f_{Td}^{(p)}=0.026\,,~f_{Ts}^{(p)}=0.118\,,\nonumber\\
[2mm]&&f_{Tc}^{(p)}=f_{Tb}^{(p)}=f_{Tt}^{(p)}=1-\sum_{q=u,d,s}^{}f^{(p)}_{Tq}=0.836\,,\nonumber\\
[2mm]
&&f_{Tu}^{(n)}=0.014\,,~f_{Td}^{(n)}=0.036\,,~f_{Ts}^{(n)}=0.118\,,\nonumber\\
[2mm]&&f_{Tc}^{(n)}=f_{Tb}^{(n)}=f_{Tt}^{(n)}=1-\sum_{q=u,d,s}^{}f^{(n)}_{Tq}=0.832\,.
\end{eqnarray}

Remarkably, the result of Eq. (\ref{si}) predicts a cancellation point where the cross section vanishes,
\begin{eqnarray}
\sigma_{\textrm{SI}}^{}= 0 \Rightarrow \kappa_1^{}= \frac{g^4_{}m_\Sigma^{}}{16\pi m_W^{}}\left(1+\frac{m_h^2}{m_W^2}\right)\,.
\end{eqnarray}

\begin{figure*}[tbp]
  \centering
  \includegraphics[width=15cm]{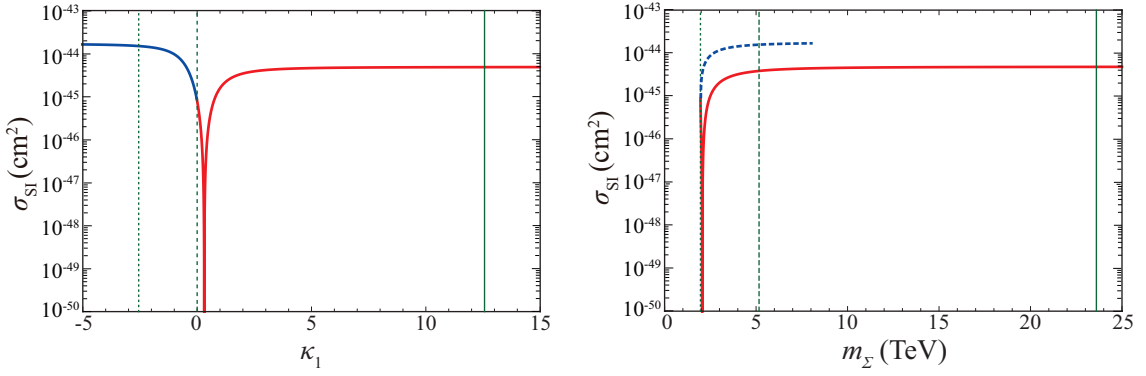}
  \caption{The dependence of the DM-nucleon scattering cross section $\sigma_{\textrm{SI}}^{}$ on the Higgs portal coupling $\kappa_1^{}$ and the DM mass $m_\Sigma^{}$. In the left panel, the dotted, dashed and solid vertical lines correspond to $\kappa_1^{}=-2.6$, $\kappa_1^{}=0$ and $\kappa_1^{}=4\pi$, respectively. In the right panel, the dashed curve is for $\kappa_1^{}<0$ and the solid curve is for $\kappa_1^{}>0$, while the dotted, dashed and solid vertical lines correspond to $m_\Sigma^{}=2\,\textrm{TeV}$, $m_\Sigma^{}=5.2\,\textrm{TeV}$ and $m_\Sigma^{}=23.6\,\textrm{TeV}$, respectively. We have $\sigma_{\textrm{SI}}^{}=1.8\times 10^{-44}_{}\,\textrm{cm}^2_{}$ for $\kappa_1^{}=-2.6$ and $m_\Sigma^{}=5.2\,\textrm{TeV}$, $\sigma_{\textrm{SI}}^{}=0.9\times 10^{-45}_{}\,\textrm{cm}^2_{}$ for $\kappa_1^{}=0$ and $m_\Sigma^{}=2\,\textrm{TeV}$, while $\sigma_{\textrm{SI}}^{}=5\times 10^{-45}_{}\,\textrm{cm}^2_{}$ for $\kappa=4\pi$ and $m_\Sigma^{}=23.6\,\textrm{TeV}$. }
  \label{SigmaPureSI}
\end{figure*}

Although the above identity would not be exactly accessible due to the $\kappa_1^{}-m_\Sigma^{}$ correlation, our result indeed exhibits an intriguing property that the spin-independent cross section $\sigma_{\textrm{SI}}^{}$ can be highly suppressed for some delicately set $\kappa_1^{}$ and $m_\Sigma^{}$, thus almost nullifying direct detection constraints on spin independent scattering. In Fig. \ref{SigmaPureSI}, we show the dependence of the DM-nucleon scattering cross section $\sigma_{\textrm{SI}}^{}$ on both the Higgs portal coupling $\kappa_1^{}$ and the DM mass $m_\Sigma^{}$. We find the coupling $\kappa_1^{}$ can significantly affect the cross section $\sigma_{\textrm{SI}}^{}$. For example, we read $\sigma_{\textrm{SI}}^{}\simeq 1.8\times 10^{-44}_{}\,\textrm{cm}^2_{}$, $0.9\times 10^{-45}\,\textrm{cm}^2_{}$ and $5\times 10^{-45}_{}\,\textrm{cm}^2_{}$ for $\kappa_1^{}=-2.6$, $0$ and $4\pi$, respectively. The cancellation manifests itself by the cross section $\sigma_{\textrm{SI}}^{}$ drastically decreasing to an extremely small value for $\kappa_1^{} \rightarrow 0.3$ or $m_\Sigma^{}\rightarrow 2.05\,\textrm{TeV}$.

\section{Dominant complex triplet}

In the dominant complex triplet scenario, the DM particle $\chi_1^0$ is mostly comprised of the real part $\delta^0_R$ of the neutral component $\delta^0_{}$ of the complex triplet $\Delta$ with hypercharge.
By taking into account the electroweak radiative corrections, the mass splittings between the DM particle and its neightbouring partners should be
\begin{eqnarray}
\label{complexsplit}
\Delta m_\Delta^{0}&=& m_{\chi^{0}_{3}}^{}-m_{\chi^{0}_{1}}^{}\simeq
m_{\delta^{0}_{I}}^{}-m_{\delta^{0}_{R}}^{} \simeq \frac{\lambda^2_{}v^4_{}}{8 m_\Sigma^{2} m_\Delta^{}}\,,\nonumber\\
[2mm]
\Delta m_\Delta^{\pm}&=& m_{\chi^{\pm}_{1}}^{}-m_{\chi^{0}_{1}}^{}\simeq m_{\delta^{\pm}_{}}^{}-m_{\delta^{0}_{R}}^{} \nonumber\\
[2mm]
&\simeq&\frac{\kappa_3^{}v^2_{}}{8m_\Delta^{}}+\frac{\lambda^2_{}v^4_{}}{16m_\Sigma^{2} m_\Delta^{}} +\frac{g^2_{}}{4\pi}m_Z^{}\left(3\sin^2_{}\frac{\theta_W^{}}{2}-2\sin^4_{}\frac{\theta_W^{}}{2}\right)\,,\nonumber\\
[2mm]
\Delta m_\Delta^{\pm\pm}&=& m_{\chi^{\pm\pm}_{}}^{}-m_{\chi^{0}_{1}}^{}\simeq m_{\delta^{\pm\pm}_{}}^{}-m_{\delta^{0}_{R}}^{} \nonumber\\
[2mm]
&\simeq&\frac{\kappa_3^{}v^2_{}}{4m_\Delta^{}}+\frac{\lambda^2_{}v^4_{}}{8m_\Sigma^{2} m_\Delta^{}}
+\frac{g^2_{}}{2\pi}m_Z^{}\sin^2_{}\theta_W^{}\,.
\end{eqnarray}
For illustration we explicitly show
\begin{eqnarray}
\Delta m_\Delta^{0}&\simeq & 176\,\textrm{MeV}\left(\frac{\lambda}{4\pi}\right)^2_{}\left(\frac{10\,m_\Delta^{}}{m_\Sigma^{}}\right)^2_{}\left(\frac{1.6\,\textrm{TeV}}{m_\Delta^{}}\right)\,,\nonumber\\
[2mm]
\Delta m_\Delta^{\pm}&\simeq &59\,\textrm{GeV} \left(\frac{\kappa_3^{}}{4\pi}\right)\left(\frac{1.6\,\textrm{TeV}}{m_\Delta^{}}\right)+88\,\textrm{MeV}\left(\frac{\lambda}{4\pi}\right)^2_{}\left(\frac{10\,m_\Delta^{}}{m_\Sigma^{}}\right)^2_{}\left(\frac{1.6\,\textrm{TeV}}{m_\Delta^{}}\right)+540\,\textrm{MeV}\,,\nonumber\\
[2mm]
\Delta m_\Delta^{\pm\pm}&\simeq & 119\,\textrm{GeV} \left(\frac{\kappa_3^{}}{4\pi}\right)\left(\frac{1.6\,\textrm{TeV}}{m_\Delta^{}}\right)+1.43\,\textrm{GeV}\,.
\end{eqnarray}

The possibility of complex triplet with hypercharge as DM has long been though excluded due to the much enhanced spin independent cross section mediated through a $Z$ boson exchange. However, the mass splitting renders the $\delta^0_I$ state heavier. Provided this mass splitting exceeds the DM kinetic energy of a few 100 keV, the inelastic scattering into the heavier state will be kinematically suppressed, thus reviving the role of DM arising from a complex triplet.

With particular interest, we can expect $\Delta m_\Delta^{\pm},\Delta m_\Delta^{\pm\pm}<m_{\pi^\pm_{}}^{}$ and $\Delta m_\Delta^{0}<m_{\pi^0_{}}^{}$ for a small $\lambda$ as well as a small and negative $\kappa_3^{}$. As a consequence, the non-hadronic decays of the non-DM components of the complex triplet with hypercharge may be well tested at the running and future colliders.

\subsection{Relic density}

Paralleling the dominant real triplet case, in order to determine the relic density of the DM particle $\chi^0_{1}\simeq \delta^0_{R}$, we include all the following annihilation and co-annihilation channels (see the relevant Feynman diagrams in appendix \ref{ann-feyndiagram}),
\begin{eqnarray}
\label{acocomplex}
\delta^{0}_{R(I)}\delta^{0}_{R(I)}&\rightarrow& W^{+}_{}W^{-}_{}\,,~ZZ\,,~\phi^{\ast}_{}\phi^{}_{}\,;\nonumber\\
[2mm]
\delta^{0}_{R}\delta^{0}_{I}&\rightarrow& W^{+}_{}W^{-}_{}\,,~f\bar{f}\,,~\phi^{\ast}_{}\phi^{}_{}\,;\nonumber\\
[2mm]
\delta^0_{R,I}\delta^\pm_{}&\rightarrow& W^\pm_{}Z\,,~W^\pm_{}A\,,~f'\bar{f}\,,~\phi^{\ast}_{}\phi^{}_{}\,;\nonumber\\
[2mm]
\delta^0_{R,I}\delta^{\pm\pm}_{}&\rightarrow& W^{\pm}_{}W^{\pm}_{}\,;\nonumber\\
[2mm]
\delta^{+}_{}\delta^{-}_{}&\rightarrow& W^+_{}W^-_{}\,,~ZZ\,,~AA\,,~ZA\,,~f\bar{f}\,,~\phi^{\ast}_{}\phi^{}_{}\,;\nonumber\\
[2mm]
\delta^{\pm}_{}\delta^{\pm}_{}&\rightarrow& W^{\pm}_{}W^{\pm}_{}\,;\nonumber\\
[2mm]
\delta^\pm_{}\delta^{\mp\mp}_{}&\rightarrow& W^{\mp}_{}Z\,, ~W^{\mp}_{}A\,,~f'\bar{f}\,,~\phi^{\ast}_{}\phi^{}_{}\,;\nonumber\\
[2mm]
\delta^{++}_{}\delta^{--}_{}&\rightarrow& W^{+}_{}W^{-}_{}\,,~ZZ\,,~AA\,,~ZA\,,~f\bar{f}\,,~\phi^{\ast}_{}\phi^{}_{}\,.\nonumber\\
&&
\end{eqnarray}
to calculate the thermally averaged cross sections up to the $p$-wave contributions,
\begin{eqnarray}
\langle\sigma^{}_{\delta^0_{R,I} \delta^0_{R,I}}v_{\textrm{rel}}^{}\rangle&=& \frac{3g^4_{}+4g^2_{}g^{\prime 2}_{}+2g^{\prime4}_{}}{16\pi m^2_{\Delta}}\left(1-\frac{5}{x}\right) +\frac{2\kappa_2^2+2\kappa_2^{}\kappa_3^{}+\kappa_3^2}{32\pi m^2_{\Delta}}\left(1-\frac{3}{x}\right) \,, \nonumber \\
[2mm]
\langle\sigma^{}_{\delta^0_{R} \delta^0_{I}}v_{\textrm{rel}}^{}\rangle&=&\frac{30g^4_{}+41g^{\prime4}_{}}{128\pi m^2_{\Delta} x}\,, \nonumber \\
[2mm]
\langle\sigma^{}_{\delta^0_{R,I} \delta^\pm_{}} v_{\textrm{rel}}^{}\rangle&=&\frac{g^2_{}\left(g^2_{}+4g^{\prime2}_{}\right)}{32\pi m^2_{\Delta} }\left[1-\frac{5g^2_{}+80g^{\prime 2}_{}}{\left(4g^2_{}+16g^{\prime 2}_{}\right)x}\right]+\frac{\kappa_3^2}{128\pi m^2_{\Delta}}\left(1-\frac{3}{x}\right) \,,\nonumber\\
[2mm]
\langle\sigma^{}_{\delta^0_{R,I} \delta^{\pm\pm}_{}} v_{\textrm{rel}}^{}\rangle&=&\frac{g^4_{}}{16\pi m^2_{\Delta}}\left( 1-\frac{5}{x}\right)\,, \nonumber\\
[2mm]
\langle\sigma^{}_{\delta^+\delta^-}v_{\textrm{rel}}^{}\rangle&=&\frac{2g^4_{}+g^{\prime4}_{}}{8\pi m^2_{\Delta}}\left[1-\frac{160g^4_{}+39g^{\prime4}_{}}{\left(32g^4_{}+16g^{\prime4}_{}\right)x}\right]+\frac{\left(2\kappa^{}_2+\kappa^{}_3\right)^2}{64\pi m^2_{\Delta}}\left(1-\frac{3}{x}\right) \,,\nonumber\\
[2mm]
\langle\sigma^{}_{\delta^\pm_{}\delta^\pm_{}}v_{\textrm{rel}}^{}\rangle&=&0\,,\nonumber\\
[2mm]
\langle\sigma^{}_{\delta^\pm_{}\delta^{\mp\mp}_{}}v_{\textrm{rel}}^{}\rangle&=&\frac{g^2_{}\left(g^2_{}+4g^{\prime2}_{}\right)}{16\pi m^2_{\Delta}}\left[1-\frac{5g^2_{}+80g^{\prime2}_{}}{\left(4g^2_{}+16g^{\prime2}_{}\right)x}\right]+\frac{\kappa_3^2}{64\pi m^2_{\Delta}}\left(1-\frac{3}{x}\right)\,, \nonumber\\
[2mm]
\langle\sigma^{}_{\delta^{++}_{}\delta^{--}_{}}v_{\textrm{rel}}^{}\rangle&=&\frac{3g^4_{}+4g^2_{}g^{\prime2}_{}+2g^{\prime4}_{}}{16\pi m^2_{\Delta}}\left[1 -\frac{90g^4_{}+160g^2_{}g^{\prime2}_{}+39g^{\prime4}_{}}{\left(24g^4_{}+32g^2_{}g^{\prime2}_{}+16g^{\prime4}_{}\right)x}\right] \nonumber \\
[2mm]
&&+\frac{2\kappa_2^2+2\kappa^{}_2\kappa^{}_3+\kappa_3^2}{32\pi m^2_{\Delta}}\left(1-\frac{3}{x}\right)\,,
\end{eqnarray}
with the definition
\begin{eqnarray}
x\equiv \frac{m_\Delta^{}}{T}\,.
\end{eqnarray}
We then obtain the effective cross section $\langle\sigma_{\text{eff}}^{\Delta}v_{\textrm{rel}}^{}\rangle$ accordingly,
\begin{eqnarray}\label{effsecc}
\langle\sigma_{\text{eff}}^\Delta v_{\textrm{rel}}^{}\rangle&=&\frac{2}{36}\langle\sigma^{}_{\delta^0_{R,I} \delta^0_{R,I}}v_{\textrm{rel}}^{}\rangle +\frac{2}{36}\langle\sigma^{}_{\delta^0_{R} \delta^0_{I}}v_{\textrm{rel}}^{}\rangle +\frac{8}{36}\langle\sigma^{}_{\delta^0_{R,I} \delta^\pm_{}} v_{\textrm{rel}}^{}\rangle +\frac{8}{36}\langle\sigma^{}_{\delta^0_{R,I} \delta^{\pm\pm}_{}} v_{\textrm{rel}}^{}\rangle \nonumber\\
[2mm]
& &+\frac{2}{36}\langle\sigma^{}_{\delta^{\pm}_{}\delta^{\mp}_{}}v_{\textrm{rel}}^{}\rangle +\frac{2}{36}\langle\sigma^{}_{\delta^{\pm}_{}\delta^{\pm}_{}}v_{\textrm{rel}}^{}\rangle +\frac{4}{36}\langle\sigma^{}_{\delta^\pm_{}\delta^{\mp\mp}_{}}v_{\textrm{rel}}^{}\rangle +\frac{2}{36}\langle\sigma^{}_{\delta^{++}_{}\delta^{--}_{}}v_{\textrm{rel}}^{}\rangle \nonumber\\
[2mm]
&=&\frac{3g^4_{}+4g^2_{}g'^2_{}+g'^4_{}}{48\pi m^2_{\Delta}}\left[1-\frac{180g^4_{}+320g^2_{}g'^2_{}+39g^{\prime4}_{}}{\left(48g^4_{}+64g^2_{}g'^2_{}+16g'^4_{}\right)x}\right] \nonumber\\
[2mm]
& &+\frac{4\kappa_2^2+4\kappa^{}_2\kappa^{}_3+3\kappa_3^2}{384\pi m^2_{\Delta}}\left(1-\frac{3}{x}\right)\,.
\end{eqnarray}
The final DM relic density are again derived through Eqs. (\ref{boltzmanneqn}-\ref{freezeout}), where $m_\Sigma^{}$ and $\langle\sigma_{\textrm{eff}}^\Sigma v_{\textrm{rel}}^{}\rangle$ are replaced by $m_\Delta^{}$ and $\langle\sigma_{\textrm{eff}}^\Delta v_{\textrm{rel}}^{}\rangle$, respectively.

\begin{figure}[htp]
  \centering
  \includegraphics[width=8cm]{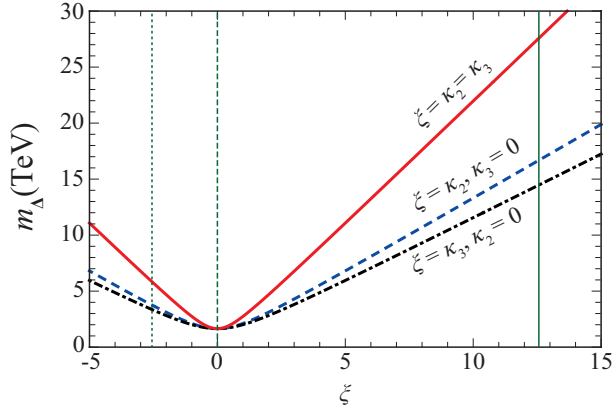}
  \caption{The correlation between the dark matter mass $m_\Delta$ and the Higgs portal couplings $(\kappa_2^{},\kappa_3^{})$. The dot-dashed, dashed and solid lines correspond to the cases with $(\kappa_2^{},\kappa_3^{})=(0,\xi)$, $(\kappa_2^{},\kappa_3^{})=(\xi,0)$ and $(\kappa_2^{},\kappa_3^{})=(\xi,\xi)$, respectively.}
  \label{DeltaPureRelic}
\end{figure}

Having at our disposal Eqs. (\ref{boltzmanneqn}-\ref{freezeout}) and (\ref{effsecc}), we can easily understand the constraint of present DM relic abundance only depends on three parameters: the mass $m_{\Delta}^{}$ of the DM scalar $\chi^0_{1}\simeq \delta^0_R$ and the two quartic couplings $\kappa_{2,3}^{}$ between the inert scalar triplet $\Delta$ and the SM Higgs doublet $\phi$. In Fig.~\ref{DeltaPureRelic}, we show the respective correlation between the DM mass $m_{\Delta}^{}$ and the Higgs portal coupling $\kappa_2^{}$, with $\kappa_3^{}=0$; and $\kappa_3^{}$, with $\kappa_2^{}=0$; as well as the case that the two couplings are set identical. For definiteness, we name a $\xi$ to be the currently concerned variable and the dot-dashed, dashed and solid curves correspond to the cases with $(\kappa_2^{},\kappa_3^{})=(0,\xi)$, $(\kappa_2^{},\kappa_3^{})=(\xi,0)$ and $(\kappa_2^{},\kappa_3^{})=(\xi,\xi)$, respectively. The two quartic couplings exhibit close behaviors in affecting the thermal relic with $\kappa_2^{}$ increasing at a faster rate, while setting them identical amounts to nearly doubling the individual effects. The lower bound of mass becomes 1.6 TeV when no quartic couplings are turned on, compared to the slightly larger value of 2 TeV in the dominant real triplet case. The upper bound, obtained at $\xi = 4\pi$ reaches $m_\Delta^{}=14.5\,\textrm{TeV}$ in the case $(\kappa_2^{},\kappa_3^{})=(0,\xi)$; $m_\Delta^{}=16.7\,\textrm{TeV}$ in the case $(\kappa_2^{},\kappa_3^{})=(\xi,0)$ and $m_\Delta^{}=27.6\,\textrm{TeV}$ in the case $(\kappa_2^{},\kappa_3^{})=(\xi,\xi)$.

As for the features remarked in the real triplet part, annihilations into SM fermions are again $p$-wave suppressed and most notable among others, including co-annihilations actually render the quantity $\langle\sigma_{\text{eff}}^\Delta v_{\textrm{rel}}^{}\rangle$ smaller roughly by a factor of 3 in the gauge interactions and 6 in the quartic interactions.

\begin{figure*}[tbp]
  \centering
  \includegraphics[width=15cm]{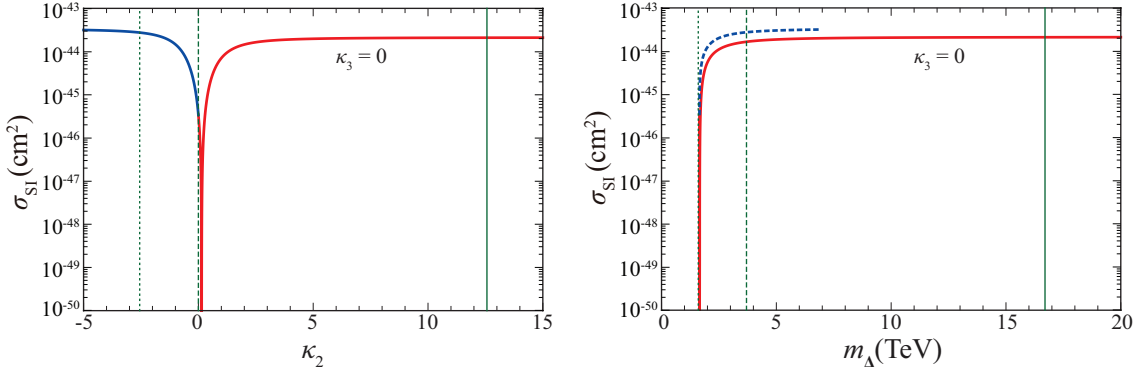}
  \caption{The dependence of the DM-nucleon scattering cross section $\sigma_{\textrm{SI}}^{}$ on the Higgs portal couplings $\kappa_2^{}$ and the DM mass $m_\Delta^{}$. In the left panels, the dotted, dashed and solid vertical lines correspond to $\kappa_2^{}=-2.6$, $\kappa_2^{}=0$ and $\kappa_2^{}=4\pi$, respectively. In the right panels, the dashed curve is for $\kappa_2^{}<0$ and the solid curve is for $\kappa_2^{}>0$, while the dotted, dashed and solid vertical lines correspond to $m_\Delta^{}=1.6\,\textrm{TeV}$, $m_\Delta^{}=3.8\,\textrm{TeV}$ and $m_\Delta^{}=16.7\,\textrm{TeV}$, respectively. We have $\sigma_{\textrm{SI}}^{}=2.7\times 10^{-44}_{}\,\textrm{cm}^2_{}$ for $\kappa_2^{}=-2.6$ and $m_\Delta^{}=3.8\,\textrm{TeV}$, $\sigma_{\textrm{SI}}^{}=3.4\times 10^{-46}_{}\,\textrm{cm}^2_{}$ for $\kappa_2^{}=0$ and $m_\Delta^{}=1.6\,\textrm{TeV}$, while $\sigma_{\textrm{SI}}^{}=2\times 10^{-44}_{}\,\textrm{cm}^2_{}$ for $\kappa_2^{}=4\pi$ and $m_\Delta^{}=16.7\,\textrm{TeV}$. }
  \label{DeltaPureSI}
\end{figure*}

\subsection{Direct detection}

The effective couplings between the DM and the quarks are given by
\begin{eqnarray}
\mathcal{L}&\supset&\frac{g^4_{}}{64\pi}\frac{m_{\Delta}^{}}{m_W^{}}\left\{\left[\frac{1}{m_W^2}+\frac{1}{m_h^2}\left(1+\frac{2}{\cos^3_{}\theta_W^{}}-\frac{64\pi }{g^4_{}} \frac{m_W^{}}{m_\Delta^{}}\kappa_2^{}\right)\right]\sum_{q}^{}m_q^{}\bar{q}q \right.\nonumber\\
[2mm]&&\left.+\frac{4}{m_W^{2}\cos\theta_W^{}}\sum_{q}^{}g_q^2 m_q^{}\bar{q}q\right\}\delta^0_{R}\delta^0_{R}~~\textrm{for}~~m_\Delta^{} \gg m_W^{} \gg m_q^{}\,,
\end{eqnarray}
with $g_{q}^{}=\frac{1}{2}-\frac{4}{3}\sin^2_{}\theta_W^{}\simeq 0.192$ for the up-type quarks $(u,c,t)$ while $g_q^{}=-\frac{1}{2}+\frac{2}{3}\sin^2_{}\theta_W^{}\simeq -0.346$ for the down-type quarks $(d,s,b)$. For the relevant Feynman diagrams, see appendix \ref{ann-feyndiagram}. The spin-independent DM-nucleon scattering cross section can be computed by
\begin{eqnarray}
\sigma_{\textrm{SI}}^{}&=&\frac{g^8_{}}{2^{12}_{}\pi^3_{}} \frac{f_N^2 m_N^4}{m_W^2} \left[\frac{1}{m_W^2}\left(1+\frac{4}{\cos\theta_W^{}}\frac{f'^{}_N}{f_N^{}}\right)+\frac{1}{m_h^2}\left(1+\frac{2}{\cos^3_{}\theta_W^{}}-\frac{64\pi }{g^4_{}} \frac{m_W^{}}{m_\Delta^{}}\kappa_2^{}\right)
\right]^2\nonumber\\
[2mm]&&\textrm{for}~~m_\Delta^{} \gg m_W^{} \gg m_N^{}\,.
\end{eqnarray}
Here the new parameter $f'^{}_N$, in analogy to the definition and calculation of the known $f_N^{}$ in Eq. (\ref{formfactor}), can be fixed by
\begin{eqnarray}
f'^{}_N&=&\frac{1}{m_N^{}}\langle N|\sum_{q}^{}g_q^2 m_q^{}\bar{q}q|N\rangle=\sum_{q=u,d,s}^{}g_q^2 f^{(N)}_{Tq}+\frac{2}{27}\sum_{q=c,b,t}^{}g_q^2 f^{(N)}_{Tq}\simeq 0.03\,.
\end{eqnarray}

In the case of dominant complex triplet, only one of the quartic couplings $\kappa_2^{}$ contributes to $\sigma_\textrm{SI}^{}$. Therefore we have displayed in Fig. \ref{DeltaPureSI} the dependence of the DM-nucleon scattering cross section $\sigma_{\textrm{SI}}^{}$ on the Higgs portal couplings $\kappa_2^{}$ and the DM mass $m_\Delta^{}$ that are correlated by the required thermal relic. The phenomenology is qualitatively similar to that in the dominant real triplet scenario except for one major quantitative difference: although still positive, the required quartic coupling strength at the cancellation point becomes a lot smaller. We read $\sigma_{\textrm{SI}}^{}=2.7\times 10^{-44}_{}\,\textrm{cm}^2_{}$ for $\kappa_2^{}=-2.6$ and $m_\Delta^{}=3.8\,\textrm{TeV}$, $\sigma_{\textrm{SI}}^{}=3.4\times 10^{-46}_{}\,\textrm{cm}^2_{}$ for $\kappa_2^{}=0$ and $m_\Delta^{}=1.6\,\textrm{TeV}$, while $\sigma_{\textrm{SI}}^{}=2\times 10^{-44}_{}\,\textrm{cm}^2_{}$ for $\kappa=4\pi$ and $m_\Delta^{}=16.7\,\textrm{TeV}$. The cancellation point is about $\kappa_2^{} \rightarrow 0.15$ or $m_\Sigma^{}\rightarrow 1.66\,\textrm{TeV}$.

\section{Democratic real and complex triplets}

As a matter of fact, for the majority of the parameter space in our mixed triplet model, either $\sin^2 \theta_0^{}\ll 1$ or $\cos^2 \theta_0^{}\ll 1$ in terms of the mixing angles in Table.~\ref{mass-table}. And within the two regions the behavior of DM particle well asymptotes to that of a pure real or complex triplet, respectively discussed in the previous two sections. The significance of relatively large mixing effects become pronounced only when the the two masses are nearly degenerate, as would be epitomized here in the democratic real and complex triplet scenario defined by Eq. (\ref{realcomplex}). In this scenario, the neutral component $\sigma^0_{}$ of the real triplet $\Sigma$ without hypercharge and the real part $\delta^0_R$ of the neutral component $\delta^0_{}$ of the complex triplet $\Delta$ with hypercharge equally constitute the DM particle $\chi^0_1$.

\subsection{Relic density}

 Apart from the annihilations and co-annihilations only involving the real or complex triplet of their own, i.e. the processes listed in Eqs. (\ref{acoreal}) and (\ref{acocomplex}), we also need to consider the following processes simultaneously cantaining the real and complex triplets to determine the DM relic density,
\begin{eqnarray}
\label{acorealcomplex}
\sigma^0_{}\delta^0_{R,I}&\longrightarrow&\phi^0_{}\phi^0_{}\,,~\phi^{0\ast}_{}\phi^{0\ast}_{}\,,\nonumber\\
[2mm]
\sigma^0_{}\delta^{\pm\pm}_{}&\longrightarrow&\phi^{\pm}_{}\phi^{\pm}_{}\,,\nonumber\\
[2mm]
\sigma^{+}_{}\delta^{-}_{}(\sigma^{-}_{}\delta^{+}_{})&\longrightarrow&\phi^{0}_{}\phi^{0}_{}(\phi^{0\ast}_{}\phi^{0\ast}_{})\,,\nonumber\\
[2mm]
\sigma^{\pm}_{}\delta^{\pm}_{}&\longrightarrow&\phi^{\pm}_{}\phi^{\pm}_{}\,,\nonumber\\
[2mm]
\sigma^{+}_{}\delta^{--}_{}(\sigma^{-}_{}\delta^{++}_{})&\longrightarrow&\phi^{0}_{}\phi^{-}_{}(\phi^{0\ast}_{}\phi^{+}_{})\,,\nonumber\\
[2mm]
\sigma^{+}_{}\delta^{0}_{R,I}(\sigma^{-}_{}\delta^{0}_{R,I})&\longrightarrow&\phi^{+}_{}\phi^{0\ast}_{}(\phi^{-}_{}\phi^{0}_{})\,.
\end{eqnarray}
The relevant Feynman diagrams are shown in appendix \ref{ann-feyndiagram}. We define
\begin{eqnarray}
x=\frac{m}{T}\,,
\end{eqnarray}
and then obtain the thermally averaged cross sections,
\begin{eqnarray}
&&\langle\sigma^{}_{ \sigma^0_{}\delta^0_{R,I}}v_{\textrm{rel}}^{}\rangle=\langle\sigma^{}_{\sigma^0_{}\delta^{\pm\pm}_{} }v_{\textrm{rel}}^{}\rangle=\langle\sigma^{}_{ \sigma^\pm_{}\delta^\mp_{}}v_{\textrm{rel}}^{}\rangle=\langle\sigma^{}_{\sigma^\pm_{}\delta^\pm_{} }v_{\textrm{rel}}^{}\rangle=\langle\sigma^{}_{\sigma^{\pm}_{}\delta^{\mp\mp}_{} }v_{\textrm{rel}}^{}\rangle =2\langle\sigma^{}_{\sigma^{\pm}_{}\delta^0_{R,I}}v_{\textrm{rel}}^{}\rangle\nonumber\\
[2mm]&=&\frac{\lambda^2_{}}{32\pi m^2_{}}\left(1-\frac{3}{x}\right)\,.
\end{eqnarray}
Incorporating all species, we eventually compute the effective cross section,
\begin{eqnarray}
\langle\sigma^{\Sigma\Delta}_{\textrm{eff}}v_{\textrm{rel}}^{}\rangle&=&\frac{9}{81}\langle\sigma^{\Sigma}_{\textrm{eff}}v_{\textrm{rel}}^{}\rangle\left|_{m_\Sigma^{}=m}^{}\right. +\frac{36}{81}\langle\sigma^{\Delta}_{\textrm{eff}}v_{\textrm{rel}}^{}\rangle\left|_{m_\Delta^{}=m}^{}\right.+\frac{1}{81}\left[4\langle\sigma^{}_{ \sigma^0_{}\delta^0_{R,I}}v_{\textrm{rel}}^{}\rangle+4\langle\sigma^{}_{\sigma^0_{}\delta^{\pm\pm}_{} }v_{\textrm{rel}}^{}\rangle\right.\nonumber\\
[2mm]
&&
\left.+2\langle\sigma^{}_{ \sigma^\pm_{}\delta^\mp_{}}v_{\textrm{rel}}^{}\rangle+2\langle\sigma^{}_{\sigma^\pm_{}\delta^\pm_{} }v_{\textrm{rel}}^{}\rangle+4\langle\sigma^{}_{\sigma^{\pm}_{}\delta^{\mp\mp}_{} }v_{\textrm{rel}}^{}\rangle+8\langle\sigma^{}_{\sigma^{\pm}_{}\delta^0_{R,I}}v_{\textrm{rel}}^{}\rangle\right]
\nonumber\\
[2mm]&=&\frac{9 g^4_{}+8g^2_{}g'^2_{}+2g'^4}{216\pi m^2_{}}\left[1 -\frac{270g^4_{}+320 g^2_{}g'^2_{}+39g^4_{}}{8(9g^4_{}+8g^2_{}g'^2_{}+2g'^4_{})x}\right]\nonumber\\
[2mm]&&+\frac{8\lambda^2_{}+2\kappa_1^2+4\kappa_2^2+4\kappa_2^{}\kappa_3^{}+3\kappa_3^2}{864 \pi m^2_{}} \left(1-\frac{3}{x}\right)\,,
\end{eqnarray}
and then solve the final DM relic density through Eqs. (\ref{boltzmanneqn}-\ref{freezeout}), where $m_\Sigma^{}$ and $\langle\sigma_{\textrm{eff}}^\Sigma v_{\textrm{rel}}^{}\rangle$ are respectively replaced by $m$ and $\langle\sigma_{\textrm{eff}}^{\Sigma\Delta} v_{\textrm{rel}}^{}\rangle$.

\begin{figure}[tbp]
  \centering
  \includegraphics[width=8cm]{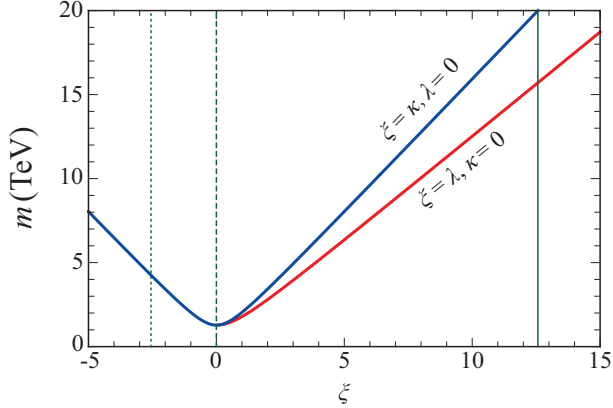}
  \caption{The correlation between the DM mass $m$ and the Higgs portal couplings $\kappa_1^{}=\kappa_2^{}=\kappa_3^{}=\kappa$ and the mixing strength $\lambda$. The upper and lower lines correspond to the cases with $(\kappa,\lambda)=(\xi,0)$ and $(\kappa,\lambda)=(0,\xi)$, respectively.}
\label{DeltaPlusSigmaRelic}
\end{figure}

 For the current concern, the DM relic density depends on the DM mass $m$ as well as the Higgs portal couplings $\kappa_{1,2,3}^{}$ and the mixing strength $\lambda$. This indicates that to accommodate the observed DM relic abundance, the DM mass $m$ should be solved by a function of these couplings. For illustration, we set the simplification $\kappa_{1}^{}=\kappa_{2}^{}=\kappa_{3}^{}=\kappa$ and plot in Fig.~\ref{DeltaPlusSigmaRelic} the correlation between $m$ and $(\kappa,\lambda)$ respectively by reducing $(\kappa,\lambda)=(\xi,0)$ and $(\kappa,\lambda)=(0,\xi)$. The situation is analogous to the pure triplet dominant cases. However, several quantitative features have altered. A notable difference is that the correct abundance can be achieved for a smaller DM mass around 1.3 TeV due to the large mixing angles and efficient co-annihilations. And the increase of mass is more gradual compared to that of $\kappa_1^{}$ in Fig.~\ref{SigmaPureRelic} and $\xi=\kappa_2^{}=\kappa_3^{}$ in Fig.~\ref{DeltaPureRelic}. The upper bound of mass, obtained at $\xi = 4\pi$ reaches $m=16\,\textrm{TeV}$ in the case $(\kappa,\lambda)=(0,\xi)$ and $m=20\,\textrm{TeV}$ in the case $(\kappa,\lambda)=(\xi,0)$.

\subsection{Direct detection}

The effective Lagrangian is composed of a coherent summation of not only the individual real and complex triplet result, but also an extra scattering amplitude in between them (see the relevant Feynman diagrams in appendix \ref{ann-feyndiagram}),
\begin{eqnarray}
\mathcal{L}&\supset&\frac{g^4_{}}{128\pi}\frac{m}{m_W^{}}\left\{\left\{\frac{5}{m_W^2} +\frac{1}{m_h^2}\left[5+\frac{2}{\cos^3_{}\theta_W^{}}-\frac{64\pi }{g^4_{}} \frac{m_W^{}}{m}(\kappa_1^{}+\kappa_2^{}-2\lambda)\right]\right\}\right. \nonumber\\
[2mm]&&\left. \times \sum_{q}^{}m_q^{}\bar{q}q+ \frac{4}{m_W^{2}\cos\theta_W^{}}\sum_{q}^{}g_q^2 m_q^{}\bar{q}q\right\}\chi^0_{1}\chi^0_{1}~~\textrm{for}~~m \gg m_W^{} \gg m_q^{}\,,
\end{eqnarray}
which yields the spin-independent DM-nucleon scattering cross section
\begin{eqnarray}
\!\!\sigma_{\textrm{SI}}^{}\!&=&\!\frac{g^8_{}}{2^{14}_{}\pi^3_{}} \frac{f_N^2 m_N^4}{m_W^2} \left\{\frac{1}{m_W^2}\left(5+\frac{4}{\cos\theta_W^{}}\frac{f'^{}_N}{f_N^{}}\right)\right.\nonumber\\
[2mm]&&\left.+\frac{1}{m_h^2}\!\left[5+\frac{2}{\cos^3_{}\theta_W^{}}-\frac{64\pi }{g^4_{}} \frac{m_W^{}}{m}(\kappa_1^{}+\kappa_2^{}-2\lambda)\right]
\!\right\}^2~~\textrm{for}~~m\gg m_W^{} \gg m_N^{}\,.
\end{eqnarray}

With the same simplification $\kappa_{1}^{}=\kappa_2^{}=\kappa_3^{}=\kappa$, we have depicted the DM-nucleon scattering cross section $\sigma_{\textrm{SI}}^{}$ as a function of both the Higgs portal coupling $\kappa$ and the mixing strength $\lambda$, accompanied by the dependence on the correlated DM mass in each case. We find close behavior to that in Fig. \ref{DeltaPureSI} with approximately the same cancellation point around $\kappa \rightarrow 0.16$ and we read $\sigma_{\textrm{SI}}^{}=2.5\times 10^{-44}_{}\,\textrm{cm}^2_{}$ for $\kappa = -2.6$ and $m = 4.2\,\textrm{TeV}$, $\sigma_{\textrm{SI}}^{}=7.1\times 10^{-46}_{}\,\textrm{cm}^2_{}$ for $\kappa = 0$ and $m = 1.3\,\textrm{TeV}$, while $\sigma_{\textrm{SI}}^{}=1.2\times 10^{-44}_{}\,\textrm{cm}^2_{}$ for $\kappa = 4\pi$ and $m = 20\,\textrm{TeV}$. The cancellation does not occur for positive $\lambda$, and we read the upper bound $\sigma_{\textrm{SI}}^{}=4\times 10^{-44}_{}\,\textrm{cm}^2_{}$ for $\lambda = 4\pi$ and $m = 15.7\,\textrm{TeV}$.

\begin{figure*}[tbp]
  \centering
  \includegraphics[width=15cm]{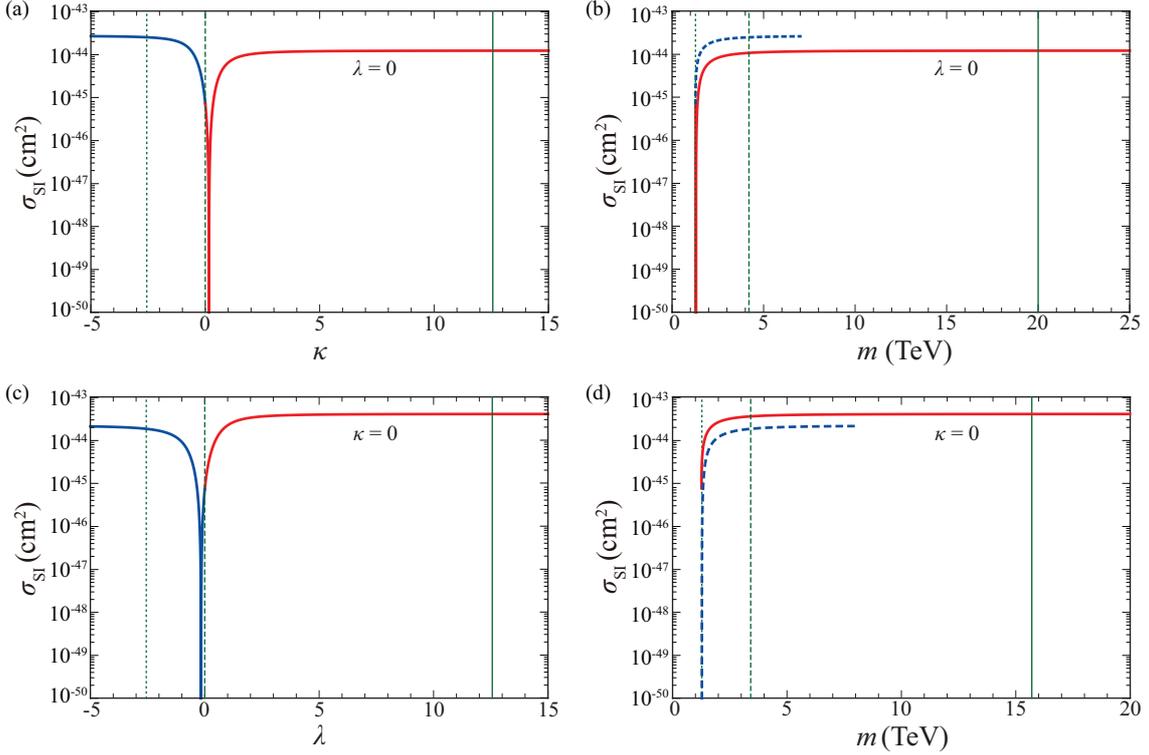}
  \caption{The upper row (panels (a) and (b)) is the dependence of DM-nucleon scattering cross section $\sigma_{\textrm{SI}}^{}$ on the Higgs portal couplings $\kappa$ and the corresponding DM mass with $\lambda =0$, and the lower row (panels (c) and (d)) is the counterpart dependence on the mixing strength $\lambda$ and the corresponding DM mass with $\kappa = 0$. In the left panel of the upper row, the dotted, dashed and solid vertical lines correspond to the coupling strength $\kappa = -2.6$, $\kappa = 0$ and $\kappa = 4\pi$, respectively, while in the right panel, the dashed curve is for $\kappa < 0$ and the solid is for $\kappa > 0$ with the vertical lines corresponding to $m =1.3\,\textrm{TeV}$, $m =4.2\,\textrm{TeV}$ and $m = 20\,\textrm{TeV}$, respectively. We have $\sigma_{\textrm{SI}}^{}=2.5\times 10^{-44}_{}\,\textrm{cm}^2_{}$ for $\kappa = -2.6$ and $m = 4.2\,\textrm{TeV}$, $\sigma_{\textrm{SI}}^{}=7.1\times 10^{-46}_{}\,\textrm{cm}^2_{}$ for $\kappa = 0$ and $m = 1.28\,\textrm{TeV}$, while $\sigma_{\textrm{SI}}^{}=1.2\times 10^{-44}_{}\,\textrm{cm}^2_{}$ for $\kappa = 4\pi$ and $m = 20\,\textrm{TeV}$. Corresponded settings apply to the lower row, and we read $\sigma_{\textrm{SI}}^{} = 1.8\times 10^{-44}_{}\,\textrm{cm}^2_{}$ for $\kappa = -2.6$ and $m = 3.4\,\textrm{TeV}$, $\sigma_{\textrm{SI}}^{}=7.1\times 10^{-46}_{}\,\textrm{cm}^2_{}$ for $\kappa = 0$ and $m = 1.28\,\textrm{TeV}$, while $\sigma_{\textrm{SI}}^{}=4\times 10^{-44}_{}\,\textrm{cm}^2_{}$ for $\kappa = 4\pi$ and $m = 15.7\,\textrm{TeV}$}
  \label{DeltaPlusSigmaSI}
\end{figure*}

\section{Radiative neutrino masses and leptogenesis}

We further introduce the following fermion doublets,
\begin{eqnarray}
X^{}_{Li}(1,2,-1/2)=\left[\begin{array}{c}
X^0_{Li}\\
[2mm]
X^{-}_{Li}
\end{array}\right]\,,~~X'^{}_{Li}(1,2,+1/2)=\left[\begin{array}{r}
-X'^{+}_{Li}\\
[2mm]
X'^{0}_{Li}
\end{array}\right]~~(i=1,...,n\geq 2)\,,
\end{eqnarray}
which take an odd parity under the $Z_2^{}$ discrete symmetry the same as the inert scalar triplets and are hence referred to as the inert fermion doublets. The Lagrangian for these inert fermions are
\begin{eqnarray}
\mathcal{L}\supset i \bar{X}^{}_L \gamma^\mu_{} D_\mu^{} X^{}_L + i \bar{X}'^{}_L \gamma^\mu_{} D_\mu^{} X'^{}_L - M_X^{} (\bar{X}^{}_L i \tau_2^{} X'^{c}_L +\textrm{H.c.})\,.
\end{eqnarray}
with the covariant derivatives
\begin{eqnarray}
D_\mu^{} X^{}_L&=&\partial_\mu^{} X^{}_L - i  \frac{1}{2} g\tau_a^{} W^a_\mu X^{}_L + i \frac{1}{2}g' B_\mu^{}X^{}_L\,, \nonumber\\
[2mm]
 D_\mu^{} X'^{}_L&=&\partial_\mu^{} X'^{}_L - i  \frac{1}{2} g\tau_a^{} W^a_\mu X'^{}_L - i \frac{1}{2}g' B_\mu^{} X'^{}_L\,.
\end{eqnarray}
The mass matrix of the inert fermions has been chosen to be real and diagonal without loss of generality and for the sake of convenience, i.e.
\begin{eqnarray}
 M_X^{} = \textrm{diag}\{M_{1}^{},~M_{2}^{},...\}\,.
 \end{eqnarray}
In this basis, we can define the Dirac fermions,
\begin{eqnarray}
X^0_i\equiv X^0_{Li}+(X'^0_{Li})^c_{}\,,~~X^{-}_i \equiv X^{-}_{Li}+(X'^{+}_{Li})^c_{}\,.
 \end{eqnarray}
with the kinetic and mass terms as below,
\begin{eqnarray}
\mathcal{L}\supset i \overline{X^0_{i}} \gamma^\mu_{} \partial_\mu^{} X^{0}_i -M_i^{}\overline{X^0_i}X_i^{0}+i \overline{X^-_{i}} \gamma^\mu_{} \partial_\mu^{} X^{-}_i -M_i^{}\overline{X^-_i}X_i^{-}\,.
\end{eqnarray}

Moreover, the inert fermion doublets and scalar triplets can have the Yukawa couplings to the SM lepton doublets,
\begin{eqnarray}
\mathcal{L}&\supset& -y\bar{l}_L^{c} i\tau_2^{} \Delta X^{}_L - y' \bar{l}_L^c i \tau_2^{} \Sigma X'^{}_L + \textrm{H.c.}\nonumber\\
[2mm]&=&-y_{\alpha i}^{}\left(-\frac{1}{\sqrt{2}}\delta^+_{}\bar{e}_{L\alpha}^c X^0_{i} - \delta^{++}_{}\bar{e}_{L\alpha}^c X^-_{i}+ \delta^0_{}\bar{\nu}_{L\alpha}^c X^0_{i}\right.\left.- \frac{1}{\sqrt{2}}\delta^+_{}\bar{\nu}_{L\alpha}^c X^-_{i}\right) \nonumber\\
[2mm]
&&-y'^{}_{\alpha i}\left(\frac{1}{\sqrt{2}}\sigma^0_{} \overline{X^{-}_{i}}e^{}_{L\alpha} - \sigma^{+}_{}\overline{X^{0}_{i}}e^{}_{L\alpha}+ \sigma^-_{} \overline{X^{-}_i} \nu_{L\alpha}^{} - \frac{1}{\sqrt{2}}\sigma^0_{} \overline{X^{0}_{i}}\nu_{L\alpha}\right)+\textrm{H.c.}\,,
\end{eqnarray}
where the SM lepton doublets are denoted by
\begin{eqnarray}
l^{}_{L\alpha}(1,2,-1/2)&=&\left[\begin{array}{c}
\nu^{}_{L\alpha}\\
[2mm]
e^{}_{L\alpha}
\end{array}\right]~~(\alpha=e,~\mu,~\tau)\,.
\end{eqnarray}

In the following we shall indicate that the inert fermions and scalars can be utilized to simultaneously generate the tiny masses of the SM neutrinos and the baryon asymmetry in the present universe.

\subsection{Radiative neutrino masses}

\begin{figure*}[tbp]
\vspace{-2cm}   \centering
  \includegraphics[width=15cm]{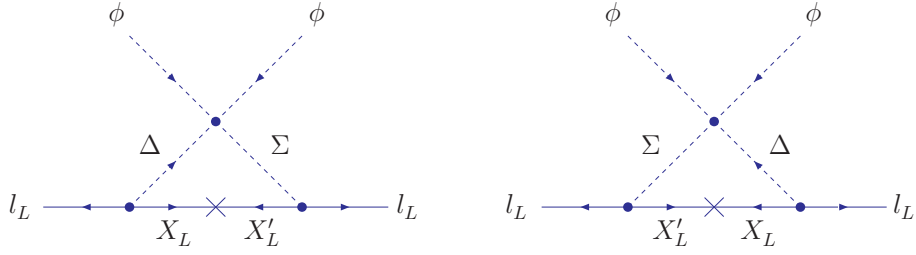}
  \vspace{-16cm}  \caption{The one-loop diagrams for generating the Majorana neutrino masses.}
  \label{numass}
\end{figure*}

As shown in Fig. \ref{numass}, the inert fermion doublets $X_{L}^{},~X'^{}_L$ and the inert scalar triplets $\Sigma,~\Delta$ can mediate a one-loop diagram to generate a Majorana mass term of the SM neutrinos $\nu_L^{}$ after the electroweak symmetry breaking,
\begin{eqnarray}
\mathcal{L}&\supset& -\frac{1}{2}m_\nu^{}\bar{\nu}_{L\alpha}^{c}\nu_{L\beta}^{}+\textrm{H.c.}\,.
\end{eqnarray}
We carry out the calculation to obtain
\begin{eqnarray}
(m_\nu^{})_{\alpha \beta}&=&\frac{\sin 2\theta_0^{}}{32\sqrt{2}\pi^2_{}}( y_{\alpha i}^{} y'^{}_{\beta i}+y'^{}_{\alpha i} y^{}_{\beta i}) M_i^{}
 \left[\frac{m_{\chi^0_1}^2}{M_i^2-m_{\chi^0_1}^2}\ln\left(\frac{M_i^2}{m_{\chi^0_1}^2}\right)-\frac{m_{\chi^0_2}^2}{M_i^2-m_{\chi^0_2}^2}\ln\left(\frac{M_i^2}{m_{\chi^0_2}^2}\right)\right]\nonumber\\
[2mm]
&&-\frac{\sin 2\theta_\pm^{}}{32\sqrt{2}\pi^2_{}} (y'^{}_{\alpha i}y^{}_{\beta i}+y_{\alpha i}^{} y'^{}_{\beta i} )M_i^{}
 \left[\frac{m_{\chi^\pm_1}^2}{M_i^2-m_{\chi^\pm_1}^2}\ln\left(\frac{M_i^2}{m_{\chi^\pm_1}^2}\right)\right.\nonumber\\
 [2mm]
&&\left.-\frac{m_{\chi^\pm_2}^2}{M_i^2-m_{\chi^\pm_2}^2}\ln\left(\frac{M_i^2}{m_{\chi^\pm_2}^2}\right)\right]\,.
\end{eqnarray}
In the limiting case with $M_i^2\gg m_{\chi^0_{1,2}}^2,~m_{\chi^\pm_{1,2}}^2 \gg \lambda v^2_{}$, the above neutrino masses can be simplified to
\begin{eqnarray}
\label{nmass}
m_\nu^{}&\simeq &\frac{\lambda v^2_{}}{16\sqrt{2}\pi^2_{}}\left( y\frac{1}{M_X^{}} y'^T_{}+ y'\frac{1}{M_X^{}} y^T_{}\right) \,.
\end{eqnarray}

It is easy to check that the neutrino masses (\ref{nmass}) can naturally arrive at a sub-eV order if the masses of the inert fermion doublets are heavy enough. For instance, we can arrange
\begin{eqnarray}
\lambda= \mathcal{O}(1)\,,~~y= y' = \mathcal{O}(1)\,,~~M_{X}^{}= \mathcal{O}(10^{13}_{}\,\textrm{GeV})\Longrightarrow m_\nu^{}=\mathcal{O}(0.1\,\textrm{eV})\,.
\end{eqnarray}
Note that for demonstration, two couples of $X^{}_L$ and $X'^{}_L$ can only produce two nonzero neutrino mass eigenvalues. If three nonzero neutrino mass eigenvalues are expected, we can introduce three or more couples of $X^{}_L$ and $X'^{}_L$.

\subsection{Leptogenesis}

\begin{figure*}[tbp]
\vspace{-4cm}   \centering
  \includegraphics[width=15cm]{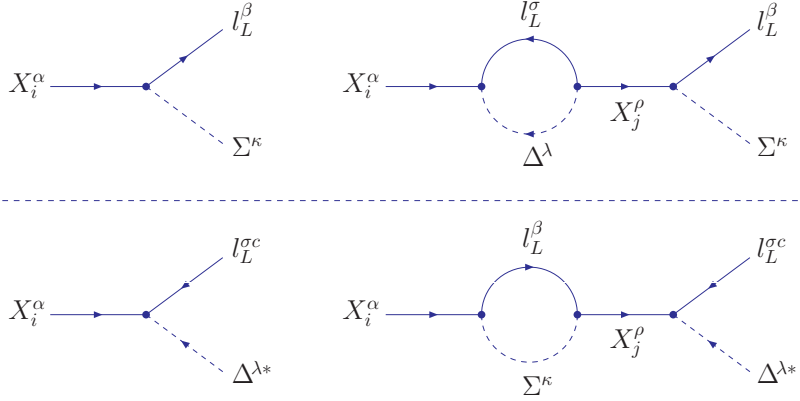}
  \vspace{-11cm}  \caption{The Dirac fermions composed of the inert fermion doublets $X_L^{}$ and $X'^{ c}_L$ decay into the SM lepton doublets $l_L^{}$ with the inert scalar triplet $\Delta$ or $\Sigma$. The CP-conjugation processes are not shown for simplicity.}
  \label{leptogenesis}
\end{figure*}

 A Dirac fermion composed of the neutral or charged components of the inert fermion doublets $X^{}_L$ and $X'^{}_L$ can decay into the SM lepton doublets $l_L^{}$ with an inert scalar triplet $\Sigma$ or $\Delta$, which is diagrammatically depicted in Fig. \ref{leptogenesis}. As a consequence of CPT-invariance and unitarity, the total and partial decay widths should respect the following relation,
\begin{eqnarray}
\Gamma_{X_i^{}}^{}&=&\Gamma(X_i^{}\longrightarrow l_L^{} \Sigma)+\Gamma(X_i^{}\longrightarrow l_L^{c} \Delta)=\Gamma(X_i^{c}\longrightarrow l_L^{c} \Sigma)+\Gamma(X_i^{c}\longrightarrow l_L^{} \Delta^\ast_{})\,.
\end{eqnarray}
Therefore, as long as the CP is not conserved, these decays can generate a lepton asymmetry stored in the SM leptons, i.e.
\begin{eqnarray}
\varepsilon_{X_i^{}}^{}&=&\frac{\Gamma(X_i^{}\longrightarrow l_L^{} \Sigma)-\Gamma(X_i^{c}\longrightarrow l_L^{c} \Sigma)}{\Gamma_i^{}}+\frac{\Gamma(X_i^{c}\longrightarrow l_L^{} \Delta^\ast_{})-\Gamma(X_i^{}\longrightarrow l_L^{c} \Delta)}{\Gamma_i^{}}\nonumber\\
[2mm]&=&2\frac{\Gamma(X_i^{}\longrightarrow l_L^{} \Sigma)-\Gamma(X_i^{c}\longrightarrow l_L^{c} \Sigma))}{\Gamma_i^{}}\neq 0\,.
\end{eqnarray}
We explicitly compute the decay width at tree level
\begin{eqnarray}
\Gamma_{X_i^{}}^{}=\frac{3}{32\pi}\left[(y^\dagger_{}y)_{ii}+y'^T_{} y'^\ast_{})_{ii}\right]M_i^{}
\end{eqnarray}
and the CP asymmetry at one-loop order,
\begin{eqnarray}
\label{cpa}
\varepsilon_{X_i^{}}^{}=\frac{3}{8\pi}\frac{\textrm{Im}[(y^\dagger_{}y)_{ji}^{}(y'^{T}_{}y'^{\ast}_{})_{ij}^{}]}{(y^\dagger_{}y)_{ii}^{}+(y'^T_{}y'^\ast_{})_{ii}^{}}\frac{M_j^{}M_i^{}}{M_j^2-M_i^2}\,.
\end{eqnarray}

For a numerical estimation, we take
\begin{eqnarray}
y'=y\,,
\end{eqnarray}
to simplify the neutrino mass matrix (\ref{nmass}) by
\begin{eqnarray}
m_\nu^{}&\simeq &\frac{\lambda }{8\sqrt{2}\pi^2_{}}y\frac{v^2_{}}{M_X^{}} y^T_{} \,,
\end{eqnarray}
and then derive an upper bound of the CP asymmetry (\ref{cpa}), i.e.
\begin{eqnarray}
|\varepsilon_{X_i^{}}^{}|<\varepsilon_{X_i^{}}^{\textrm{max}}=\frac{3\pi}{\sqrt{2}\lambda}\frac{M_{X_i^{}}^{}m_{\textrm{max}}^{}}{v^2_{}}\,.
\end{eqnarray}
Here $m_{\textrm{max}}^{}=\mathcal{O}(0.1\,\textrm{eV})$ is the maximal eigenvalue of the neutrino mass matrix. If the lightest inert fermion $X_{1}^{}$ is further assumed to be much lighter than the heavier ones $X_i^{},i\geq 2$, the final baryon asymmetry should primarily come from the $X_{1}^{}$ decays. We then define
\begin{eqnarray}
\label{rwidth}
K&=&\frac{\Gamma_{X_{1}^{}}^{}}{2H(T)}\left|_{T=M_{X_{1}^{}}^{}}^{}\right.\,,
\end{eqnarray}
where $H(T)$ is the Hubble constant,
\begin{eqnarray}
H=\left(\frac{8\pi^{3}_{}g_{\ast}^{}}{90}\right)^{\frac{1}{2}}_{}
\frac{T^{2}_{}}{M_{\textrm{Pl}}^{}}\,,
\end{eqnarray}
with $g_{\ast}^{}$ being the relativistic degrees of freedom during the leptogenesis epoch. In the strong washout region where
\begin{eqnarray}
1\ll K\lesssim  10^6_{}\,,
\end{eqnarray}
the final baryon asymmetry can well approximate to
\begin{eqnarray}
\eta_B^{}=\frac{n_B^{}}{s}\simeq -\frac{28}{79}\times \frac{\varepsilon_{X_{1}^{}}^{}}{g_\ast^{}K z_f^{}} \times 2~~\textrm{with}~~z_f^{}=\frac{M_{X_1^{}}^{}}{T_f^{}}\simeq 4.2(\ln K)^{0.6}_{}\,.
\end{eqnarray}
Here $n_B^{}$ and $s$ are respectively the baryon number density and the entropy density, the factor $-\frac{28}{79}$ is the sphaleron lepton-to-baryon coefficient, while the factor $2$ appears because the decaying particle $\eta_{1}^{}$ is a doublet. After fixing $g_\ast^{}=115.75$ (the SM fields plus one real scalar triplet and one complex scalar triplet) and setting the inputs,
\begin{eqnarray}
\label{pchoice}
M_{X_{1}^{}}^{}=0.1\,M_{X_2^{}}^{}=10^{13}_{}\,\textrm{GeV}\,,~~y=y'=\mathcal{O}(1)\,,~~\lambda=\mathcal{O}(1)\,,
\end{eqnarray}
we read
\begin{eqnarray}
K=\mathcal{O}(200)\,,~~z_f^{}=\mathcal{O}(10)\,,~~T_f^{}=\mathcal{O}(10^{12}_{}\,\textrm{GeV})\,,~~\varepsilon_{X_{1}^{}}^{\textrm{max}}=\mathcal{O}(0.1)\,.
\end{eqnarray}
The baryon asymmetry then can arrive at an expected value,
\begin{eqnarray}
\label{basy}
\eta_B^{}= 10^{-10}_{}\left(\frac{\varepsilon_{X_{1}^{}}^{}}{3.3\times10^{-5}_{}}\right)\,.
\end{eqnarray}

\section{Summary}

In this paper we have explored a mixed inert scalar triplet DM scenario where a complex scalar triplet with non-zero hypercharge and a real scalar triplet with zero hypercharge mix with each other through their renormalizable coupling to the SM Higgs doublet. By investigating the DM phenomenology regarding relic abundance and direct detection, we have systematically studied three specified regions, namely the dominant real triplet, the dominant complex triplet and the democratic real and complex triplets, which respectively correspond to the three limiting scenarios where (i) the neutral component of the real triplet dominates the DM particle, (ii) the neutral component of the complex triplet dominates the DM particle, and (iii) the neutral components of the real and complex triplets equally contribute to the DM particle. Furthermore, we have introduced two types of inert fermion doublets with opposite hypercharges to construct some heavy Dirac fermions. The Yukawa couplings of the inert scalar triplets and fermion doublets to the SM lepton doublets can be utilized to generate the Majorana neutrino masses at one-loop level and realize a successful leptogenesis for the cosmic baryon asymmetry.

\acknowledgments

The authors were supported by the Recruitment Program for Young Professionals under Grant No. 15Z127060004, the Shanghai Jiao Tong University under Grant No. WF220407201 and the Shanghai Laboratory for Particle Physics and Cosmology under Grant No. 11DZ2260700. This work was also supported by the Key Laboratory for Particle Physics, Astrophysics and Cosmology, Ministry of Education.

\appendix

\section{Couplings of the individual components of the inert real and complex scalar triplets to the standard mdoel}

For the inert real triplet $\Sigma$ without hypercharge, its components $\sigma^0_{}$ and $\sigma^{\pm}_{}$ acquire the gauge and scalar couplings to the SM fields as follows,
\begin{eqnarray}
\mathcal{L}^\Sigma_{} &\supset&g^2_{}\left[8\sigma^+_{}\sigma^-_{}(W^+_\mu W^{+\mu}_{}+W^3_\mu W^{3\mu}_{})+8\sigma^0_{}\sigma^0_{}W^+_\mu W^{-\mu}_{} \right. \nonumber\\
[2mm]
&&\left. -\left(4\sigma^{\pm}_{}\sigma^{\pm}_{}W^{\mp}_\mu W^{\mp\mu}_{}
+8 \sigma^{\pm}_{}\sigma^0_{}W^{\pm}_\mu W^{3\mu}_{}\right)\right] +i 4g \left[(\sigma^+_{}\partial_\mu^{}\sigma^0_{}-\sigma^0_{}\partial_\mu^{}\sigma^+_{})W^{-\mu}_{} \right. \nonumber\\
[2mm]
&&\left. +(\sigma^0_{}\partial_\mu^{}\sigma^-_{}-\sigma^-_{}\partial_\mu^{}\sigma^0_{})W^{+\mu}_{}
+(\sigma^+_{}\partial_\mu^{}\sigma^-_{}-\sigma^-_{}\partial_\mu^{}\sigma^+_{})W^{3\mu}_{}\right] \nonumber\\
[2mm]
&& -\frac{1}{2}\kappa_1^{}\sigma^0_{}\sigma^0_{}(\phi^{0\ast}_{}\phi^0_{}+\phi^+_{}\phi^-_{})+\kappa_1^{}\sigma^+_{}\sigma^-_{}(\phi^{0\ast}_{}\phi^0_{}+\phi^+_{}\phi^-_{})\,.
\end{eqnarray}

For the inert complex triplet $\Delta$ with hypercharge, its components $\delta^0_{}=\frac{1}{\sqrt{2}}(\delta^0_{R}+i\delta^0_I)$, $\delta^{\pm}_{}$ and $\delta^{\pm\pm}_{}$ acquire the gauge and scalar couplings to the SM fields as follows,
\begin{eqnarray}
\mathcal{L}^\Delta_{} &\supset&\delta^{0\ast}_{}\delta^0_{}\left[g^2_{} W^{+}_\mu  W^{-\mu}_{} + (g^2_{}+g'^2_{})Z^{}_\mu Z^\mu_{}\right]+\delta^{-}_{}\delta^{+}_{}\left[2g^2_{}W^{+}_\mu W^{-\mu}_{} \right. \nonumber\\
[2mm]&& \left.+ \frac{g'^2_{}}{g^2_{}+g'^2_{}}\left(g^2_{} A_\mu^{}A^\mu_{}+g'^2_{}Z_\mu^{}Z^\mu_{} -2gg'A_\mu^{}Z^\mu_{}\right) \right]+\delta^{--}_{}\delta^{++}_{}\left\{g^2_{}W^{+}_\mu W^{-\mu}_{}\right. \nonumber\\
[2mm]&&\left. +  \frac{1}{g^2_{}+g'^2_{}}(2gg'A^{}_\mu+(g^2_{}-g'^2_{}) Z_\mu^{} )(2gg'A^{}_\mu+(g^2_{}-g'^2_{})Z_\mu^{}\right\} \nonumber\\
[2mm]&& +\frac{g}{\sqrt{g^2_{}+g'^2_{}}}\left\{\delta^{-}_{}\delta^0_{}\left[gg'A_\mu^{}-(g^2_{}+2gg')Z_\mu^{}\right]W^{-\mu}_{} \right. \nonumber\\
[2mm]&&\left. + \delta^{0\ast}_{}\delta^+_{}\left[gg'A_\mu^{}-(g^2_{}+2gg')Z_\mu^{}\right]W^{+\mu}_{}\right\} \nonumber\\
[2mm]&&-\frac{g}{\sqrt{g^2_{}+g'^2_{}}}\left\{\delta^{--}_{}\delta^{+}_{}\left[3gg'A_\mu^{}+(g^2_{}-2g'^2_{})Z_\mu^{}\right]W^{+\mu}_{}\right. \nonumber\\
[2mm]&&\left.+\delta^{-}_{}\delta^{++}_{}\left[3gg'A_\mu^{}+(g^2_{}-2g'^2_{})Z_\mu^{}\right]W^{-\mu}_{} \right\}\nonumber\\
[2mm]&&-g^2_{}(\delta^{0\ast}_{}\delta^{++}_{}W^{-\mu}_{}W^{-\mu}_{}+\delta^{--}_{}\delta^0_{}W^{+\mu}_{}W^{+\mu}_{})+\sqrt{g^2_{}+g'^2_{}}(\delta^0_R\partial_\mu^{}\delta^0_I - \delta^0_I\partial_\mu^{}\delta^0_R) Z^\mu_{}\nonumber\\
[2mm]&&+i \frac{g'}{\sqrt{g^2_{}+g'^2_{}}}(\delta^{-}_{}\partial_\mu^{} \delta^{+}_{}-\delta^{+}_{}\partial_\mu^{}\delta^{-}_{})(gA^\mu_{}-g'^{}Z^\mu_{}) \nonumber\\
[2mm]&& +i \frac{1}{\sqrt{g^2_{}+g'^2_{}}}(\delta^{--}_{}\partial_\mu^{} \delta^{++}_{}-\delta^{++}_{}\partial_\mu^{}\delta^{--}_{}) [2gg'A^\mu_{}+(g^2_{}-g'^{})Z^\mu_{}] \nonumber\\
[2mm]&& + i g[(\delta^{0}_{}\partial_\mu^{} \delta^{-}_{}-\delta^{-}_{}\partial_\mu^{}\delta^{0}_{})W^{+\mu}_{}-(\delta^{0\ast}_{}\partial_\mu^{} \delta^{+}_{}-\delta^{+}_{}\partial_\mu^{}\delta^{0\ast}_{})W^{-\mu}_{}] \nonumber\\
[2mm]&&
+ i g[(\delta^{+}_{}\partial_\mu^{} \delta^{--}_{}-\delta^{--}_{}\partial_\mu^{}\delta^{+}_{})W^{+\mu}_{}-(\delta^{-}_{}\partial_\mu^{} \delta^{++}_{}-\delta^{++}_{}\partial_\mu^{}\delta^{-}_{})W^{-\mu}_{}]\nonumber\\
[2mm]&&-\delta^{0\ast}_{}\delta^0_{}[\kappa_2^{}\phi^{0\ast}_{}\phi^0_{}+(\kappa_2^{}+\kappa_3^{})\phi^{+}_{}\phi^{-}_{}]
-\frac{1}{\sqrt{2}}\kappa_3^{}\left(\delta^{-}_{}\delta^0_{}\phi^{+}_{}\phi^0_{}+\delta^{0\ast}_{}\delta^{+}_{}\phi^{0\ast}_{}\phi^{-}_{} \right. \nonumber\\
[2mm]&&
\left. +\delta^{--}_{}\delta^{+}_{}\phi^{+}_{}\phi^0_{}+\delta^{-}_{}\delta^{++}_{}\phi^{0\ast}_{}\phi^{-}_{}\right) -\left(\kappa_2^{}+\frac{1}{2}\kappa_3^{}\right)\delta^-_{}\delta^+_{}(\phi^{0\ast}_{}\phi^0_{}+\phi^{+}_{}\phi^{-}_{}) \nonumber\\
[2mm]&& -\delta^{--}_{}\delta^{++}_{}[(\kappa_2^{}+\kappa_3^{})\phi^{0\ast}_{}\phi^{0}_{})+\kappa_2^{}\phi^{+}_{}\phi^{-}_{}]\,.
\end{eqnarray}

The inert real and complex triplets $\Sigma$ and $\Delta$ also simultaneously couple to the SM fields through
\begin{eqnarray}
\mathcal{L}^{\Sigma\Delta}_{} &\supset&-\lambda\left[\frac{1}{\sqrt{2}}(\sigma^0_{}\delta^0_{}-\sigma^{-}_{}\delta^{+}_{})\phi^0_{}\phi^0_{}
+\frac{1}{\sqrt{2}}(\sigma^0_{}\delta^{++}_{}-\sigma^+_{}\delta^+_{})\phi^{-}_{}\phi^{-}_{}\right.\nonumber\\
&& \left.+(\sigma^{+}_{}\delta^0_{}-\sigma^{-}_{}\delta^{++}_{})\phi^0_{}\phi^{-}+\textrm{H.c.}\right]\,.
\end{eqnarray}

\section{Feynman diagrams for the dark matter annihilations and scattering} \label{ann-feyndiagram}

\clearpage
\newpage

\begin{figure*}[tbp]
\vspace{-5cm}   \centering
  \includegraphics[width=15cm]{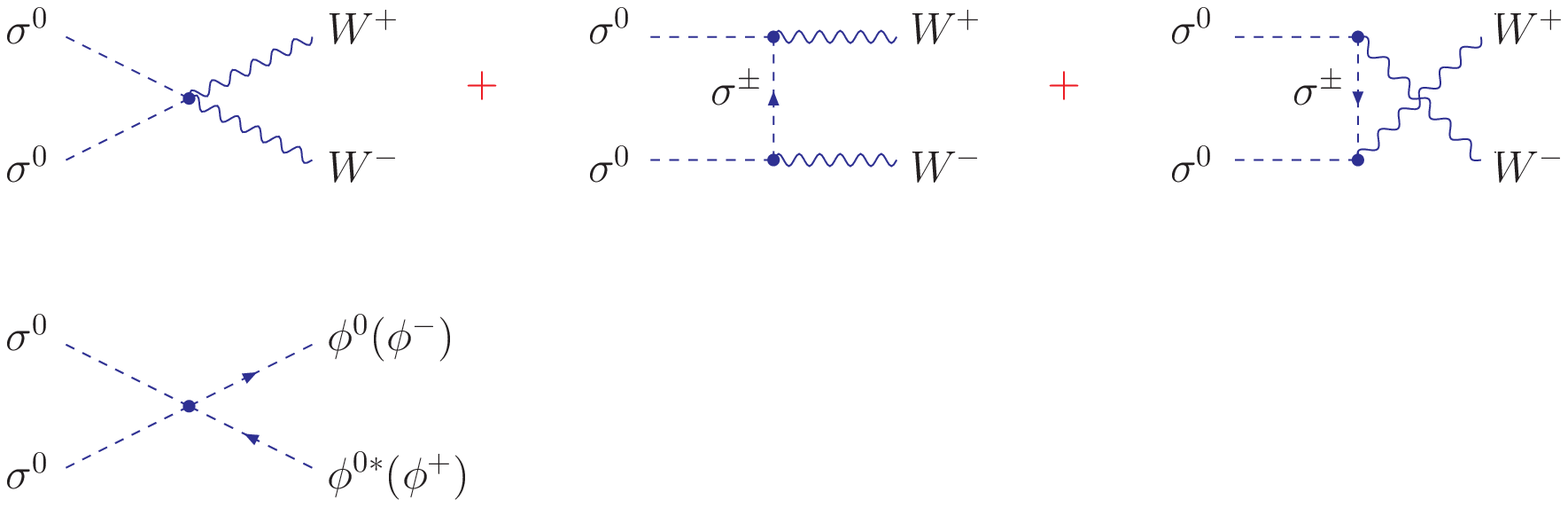}
  \vspace{-14cm}  \caption{The $\sigma^0_{}\,+\,\sigma^0_{}$ annihilations.}
  \label{s0s0}
\end{figure*}

\begin{figure*}[tbp]
\vspace{-3cm}   \centering
  \includegraphics[width=15cm]{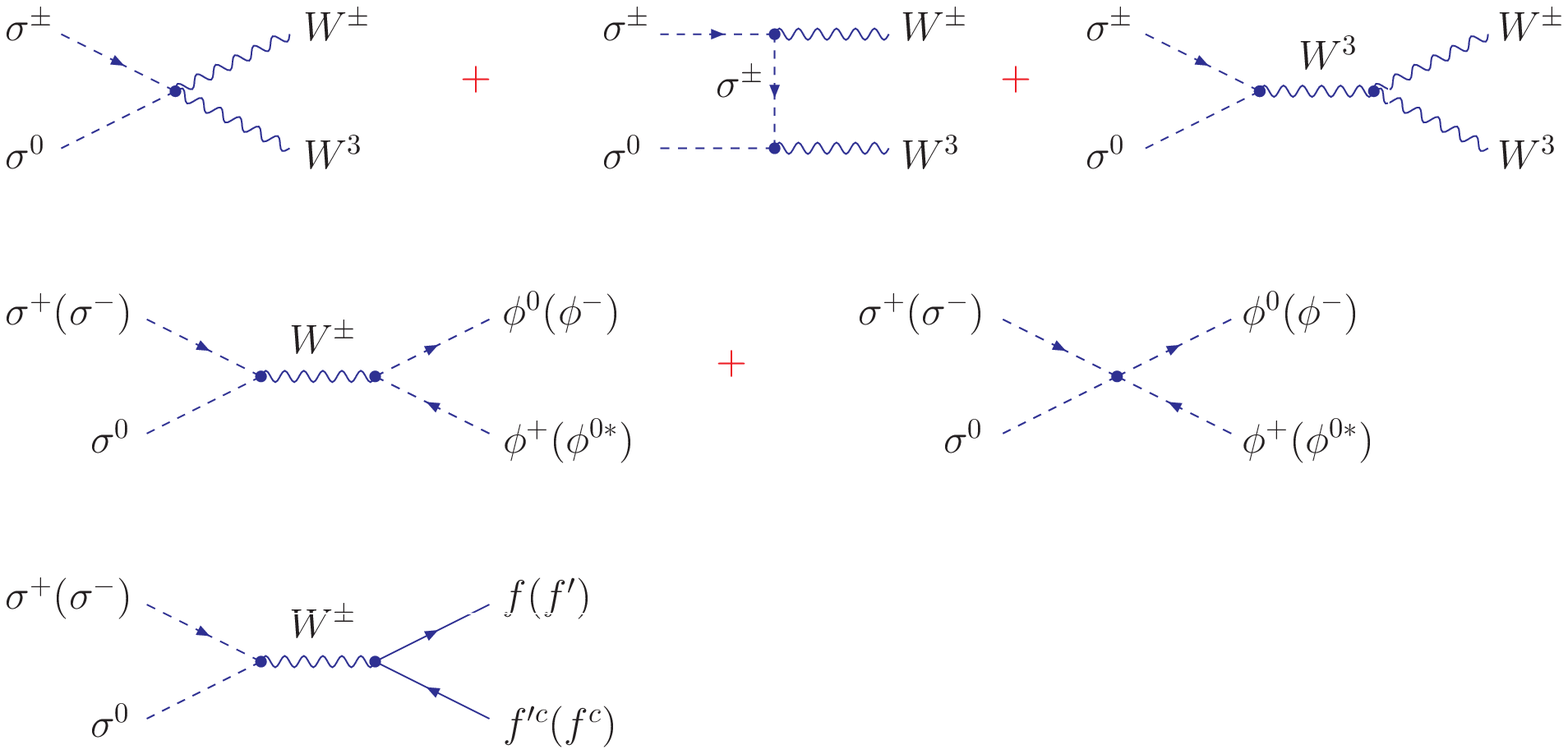}
  \vspace{-11.5cm}  \caption{The $\sigma^0_{}\,+\,\sigma^{\pm}_{}$ annihilations.}
  \label{s0s+-}
\end{figure*}

\clearpage
\newpage

\begin{figure*}[tbp]
\vspace{-3cm}   \centering
  \includegraphics[width=15cm]{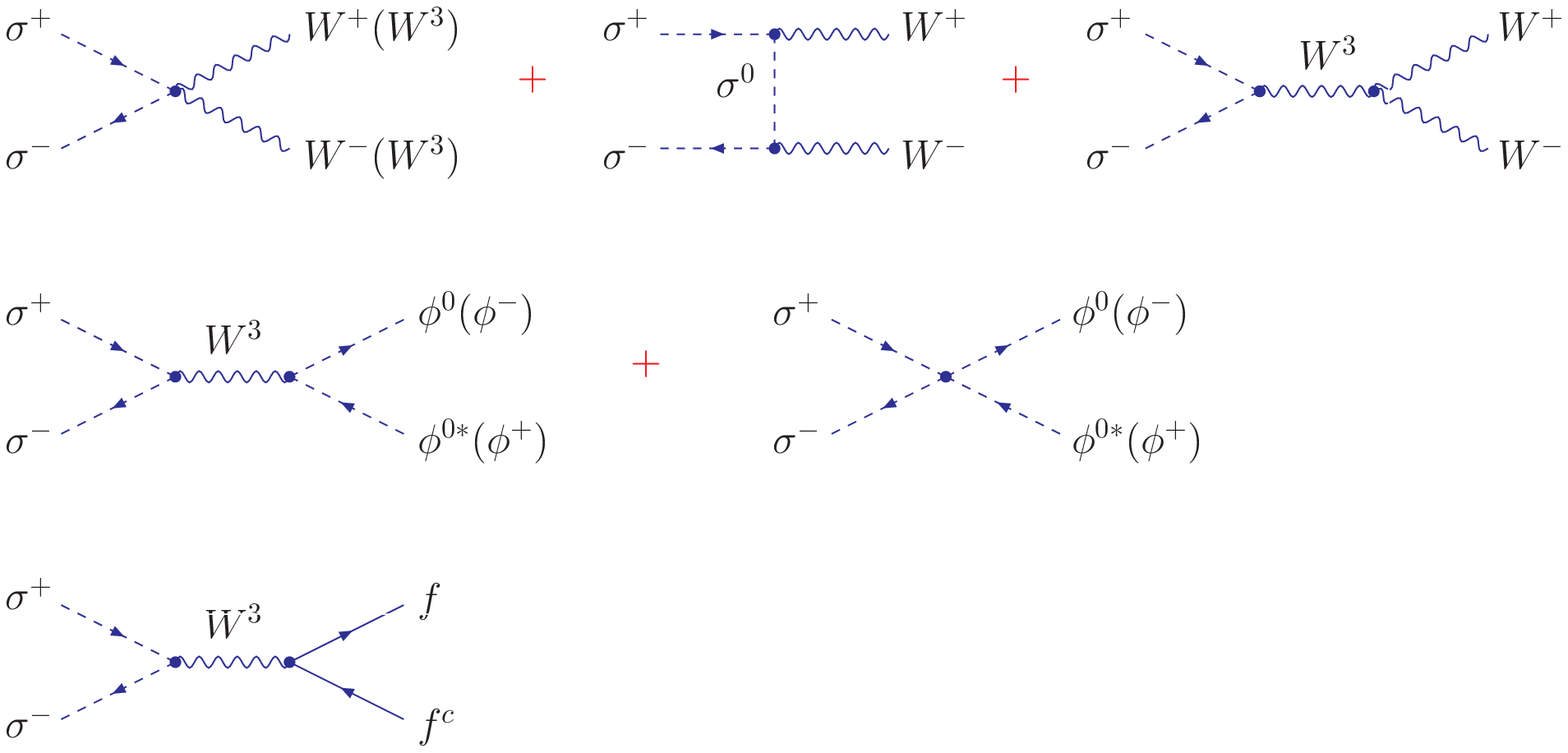}
  \vspace{-11.5cm}  \caption{The $\sigma^+_{}\,+\,\sigma^{-}_{}$ annihilations.}
  \label{s+s-}
\end{figure*}

\begin{figure*}[tbp]
\vspace{-3cm}   \centering
  \includegraphics[width=15cm]{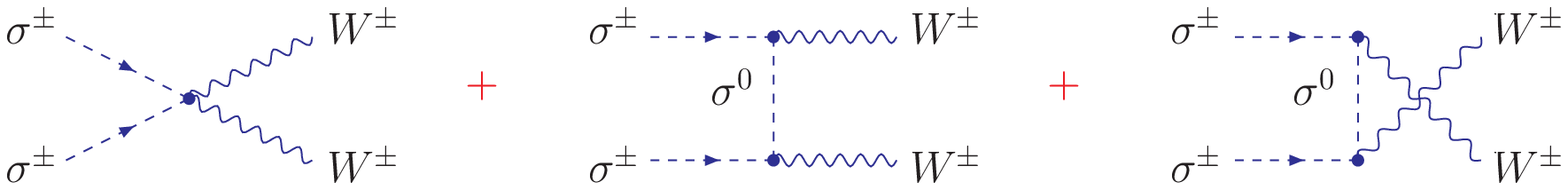}
  \vspace{-16cm}  \caption{The $\sigma^\pm_{}\,+\,\sigma^{\pm}_{}$ annihilations.}
  \label{s+-s+-}
\end{figure*}

\clearpage
\newpage

\begin{figure*}[tbp]
\vspace{-3cm}   \centering
  \includegraphics[width=15cm]{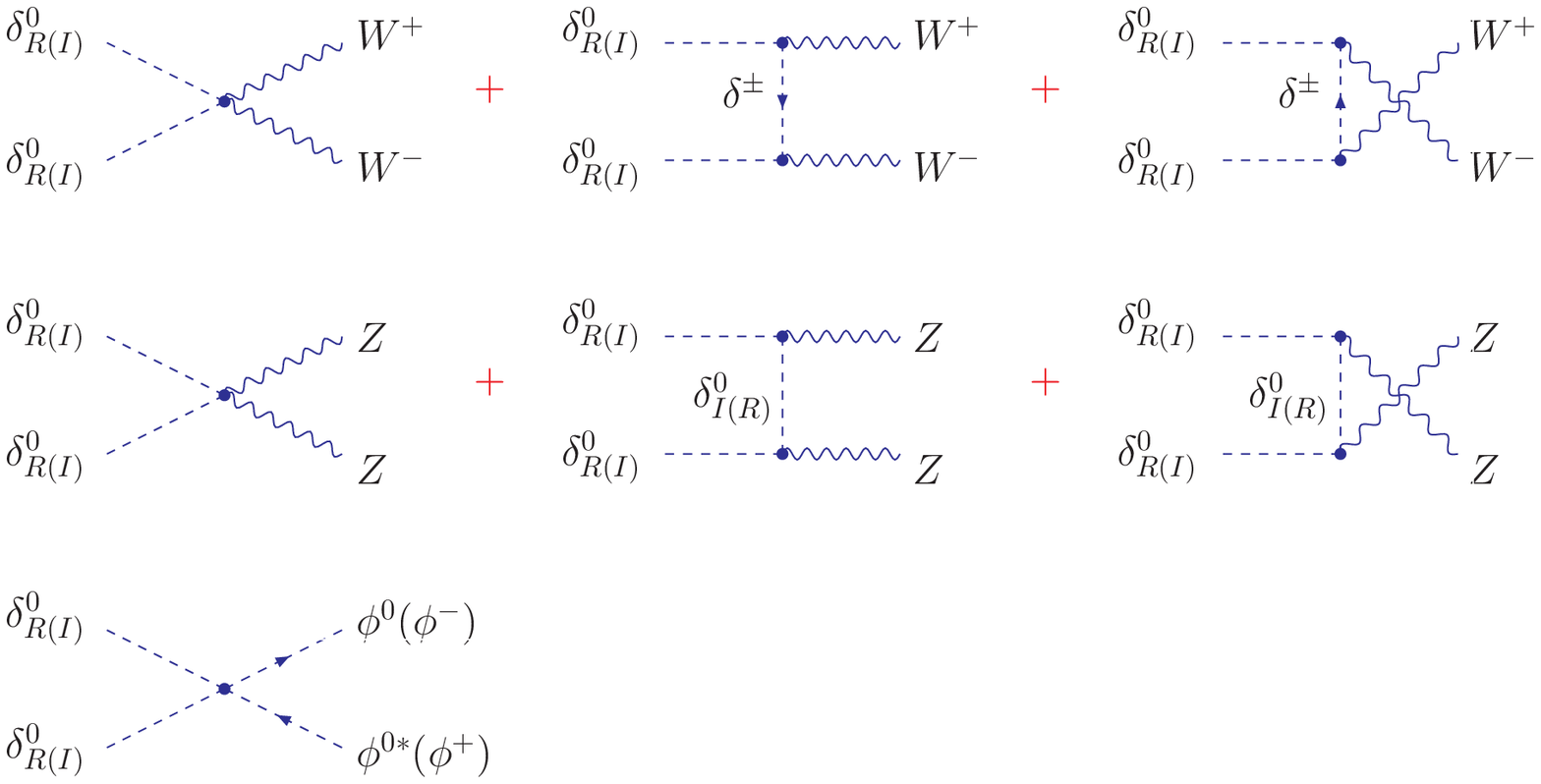}
  \vspace{-11.5cm}  \caption{\label{d0rd0r} The $\delta^0_{R(I)}\,+\,\delta^0_{R(I)}$ annihilations.}  \label{s+-s+-}
\end{figure*}

\begin{figure*}[tbp]
\vspace{-3cm}   \centering
  \includegraphics[width=15cm]{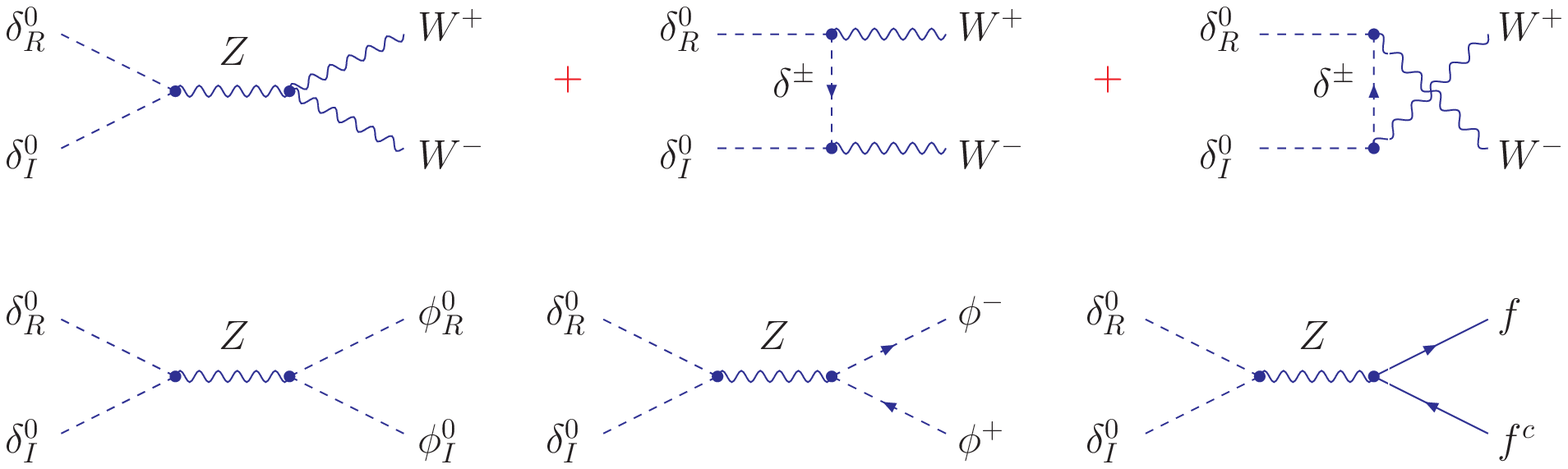}
  \vspace{-13.5cm}  \caption{\label{d0rd0i} The $\delta^0_R\,+\,\delta^0_I$ annihilations.}
\end{figure*}

\begin{figure*}[tbp]
\vspace{-3cm}   \centering
  \includegraphics[width=15cm]{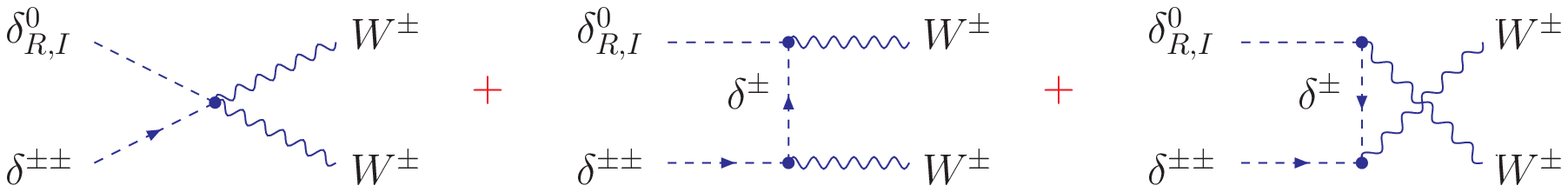}
  \vspace{-16cm}  \caption{\label{d0d++--} The $\delta^0_{R,I}\,+\,\delta^{\pm\pm}_{}$ annihilations.}
\end{figure*}

\clearpage
\newpage

\begin{figure*}[tbp]
\vspace{-3cm}   \centering
  \includegraphics[width=15cm]{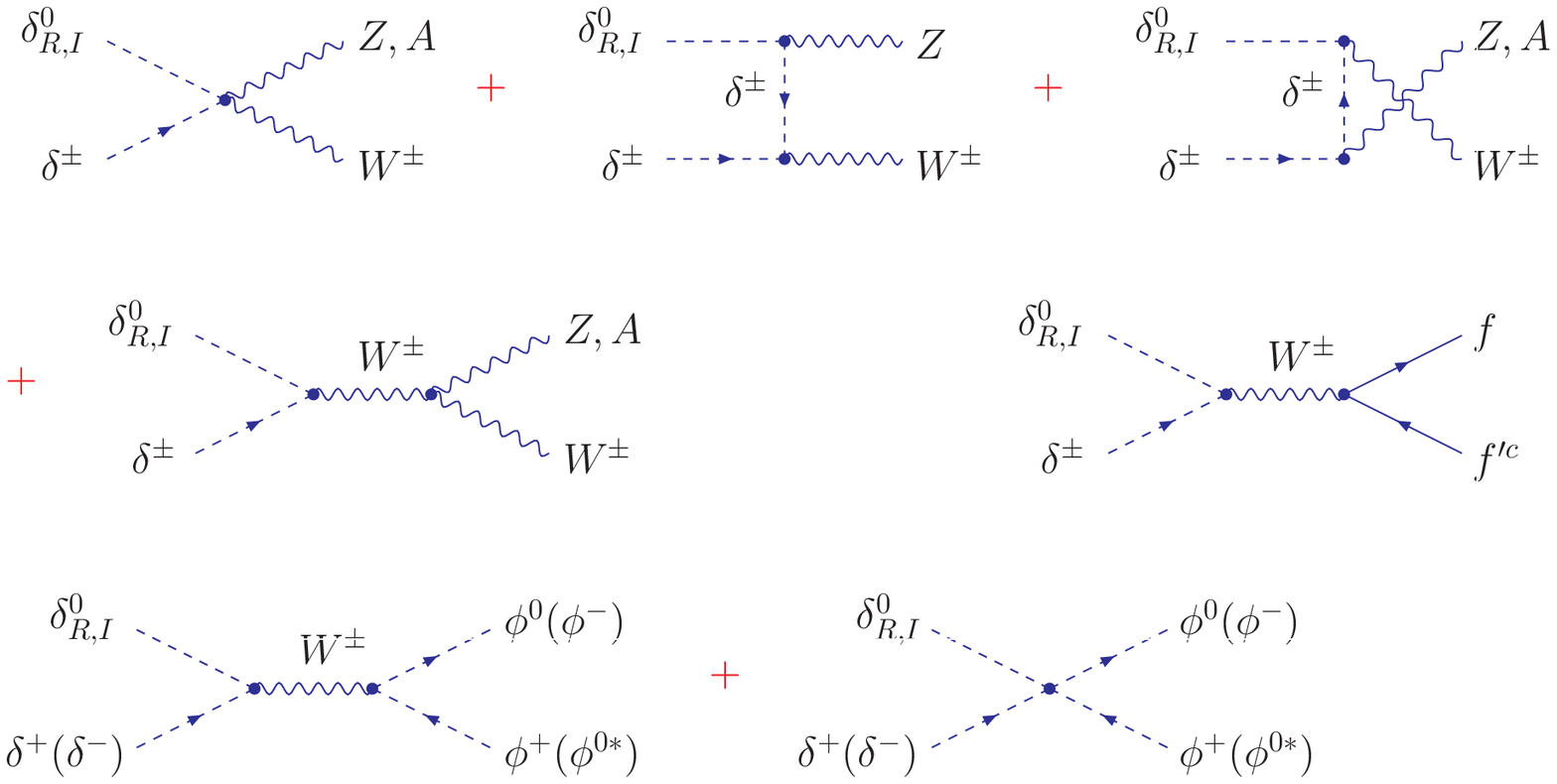}
  \vspace{-11.5cm} \caption{\label{d0d+-} The $\delta^0_{R,I}\,+\,\delta^\pm_{}$ annihilations.}
\end{figure*}

\begin{figure*}[tbp]
\vspace{-1cm}   \centering
  \includegraphics[width=15cm]{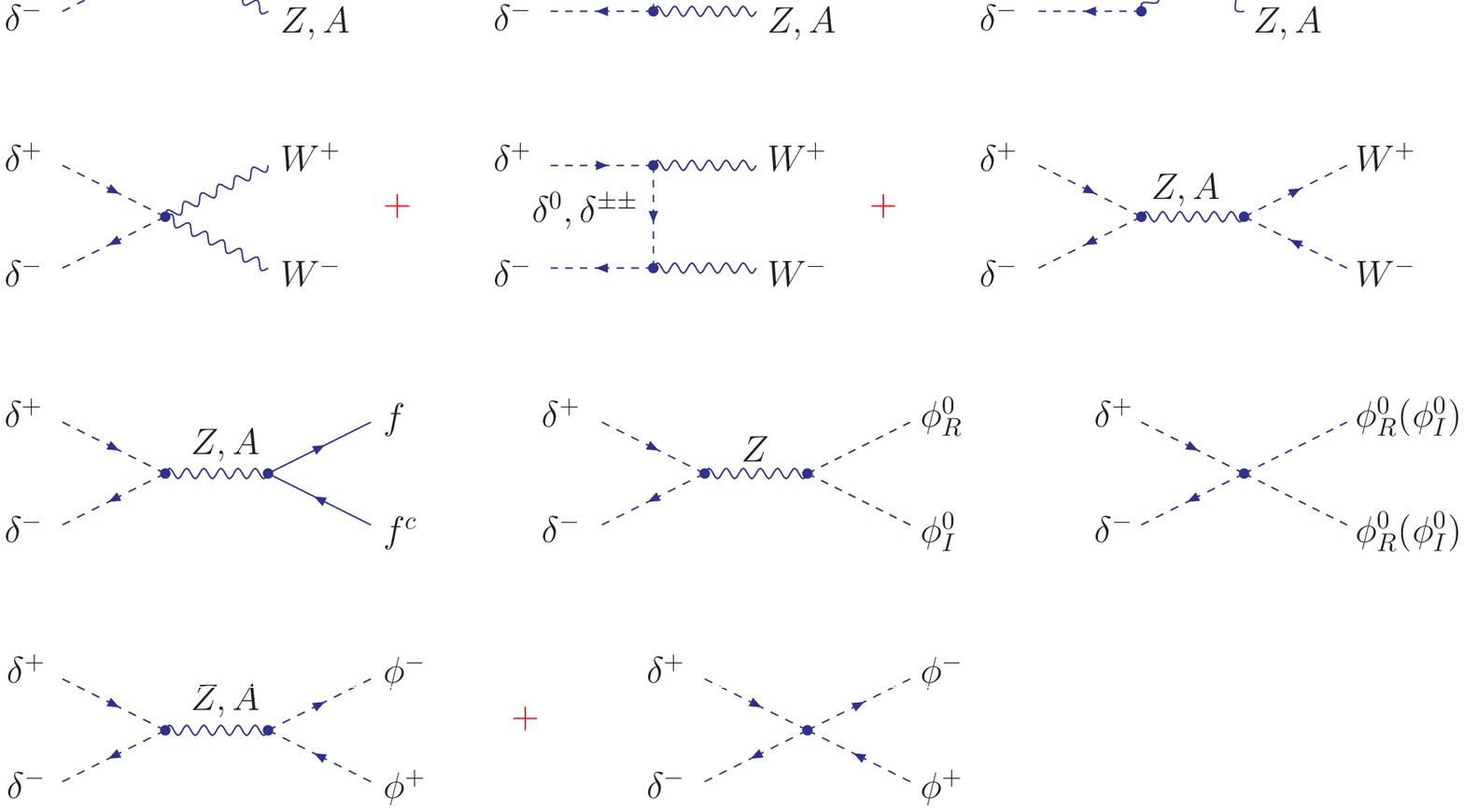}
  \vspace{-11.5cm} \caption{\label{d+d-} The $\delta^{\pm}_{}\,+\,\delta^{\mp}_{}$ annihilations.}
\end{figure*}

\clearpage
\newpage

\begin{figure*}[tbp]
\vspace{-3cm}   \centering
  \includegraphics[width=15cm]{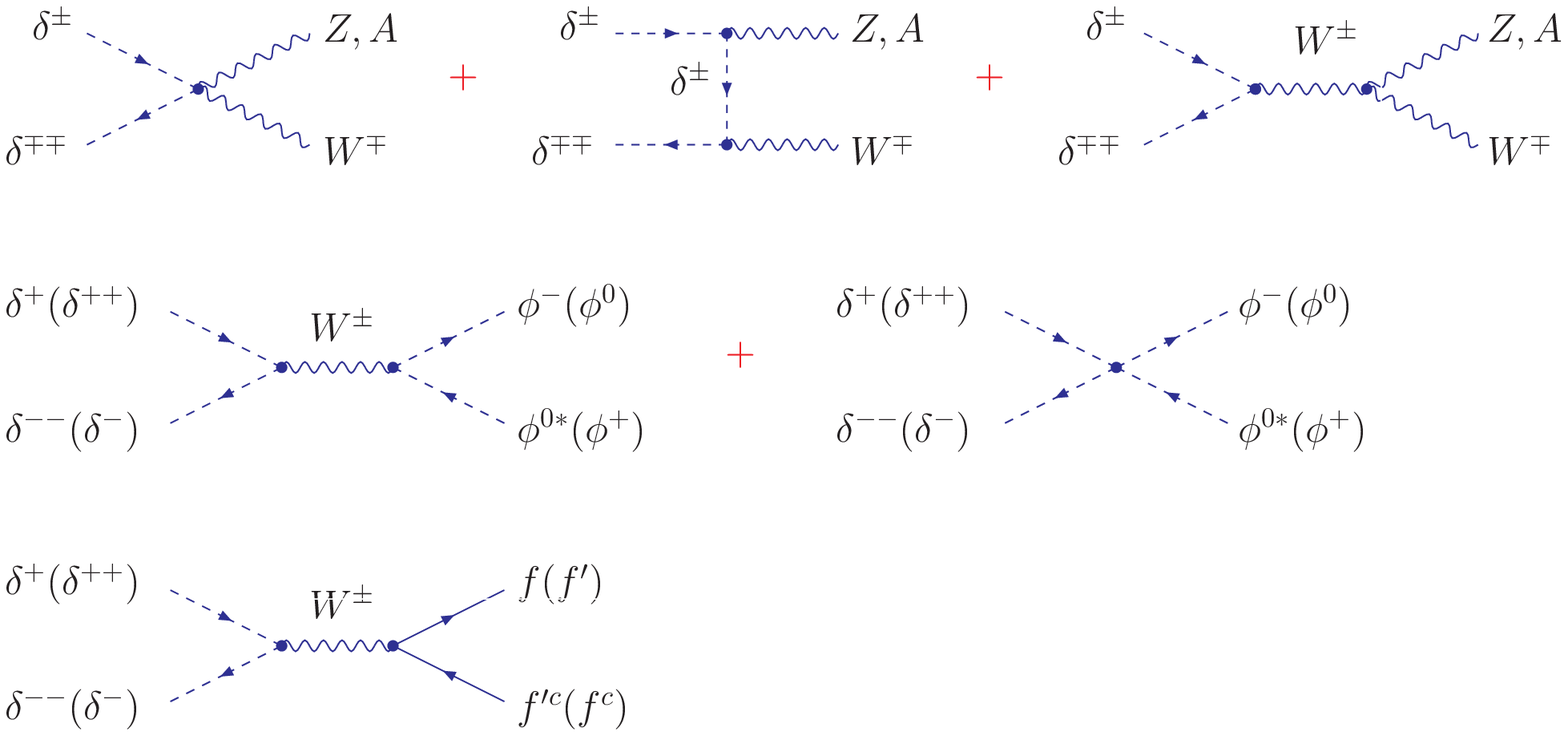}
  \vspace{-11.5cm} \caption{\label{d+-d++--} The $\delta^{\pm}_{}\,+\,\delta^{\mp\mp}_{}$ annihilations.}
\end{figure*}

\begin{figure*}[tbp]
\vspace{-1cm}   \centering
  \includegraphics[width=15cm]{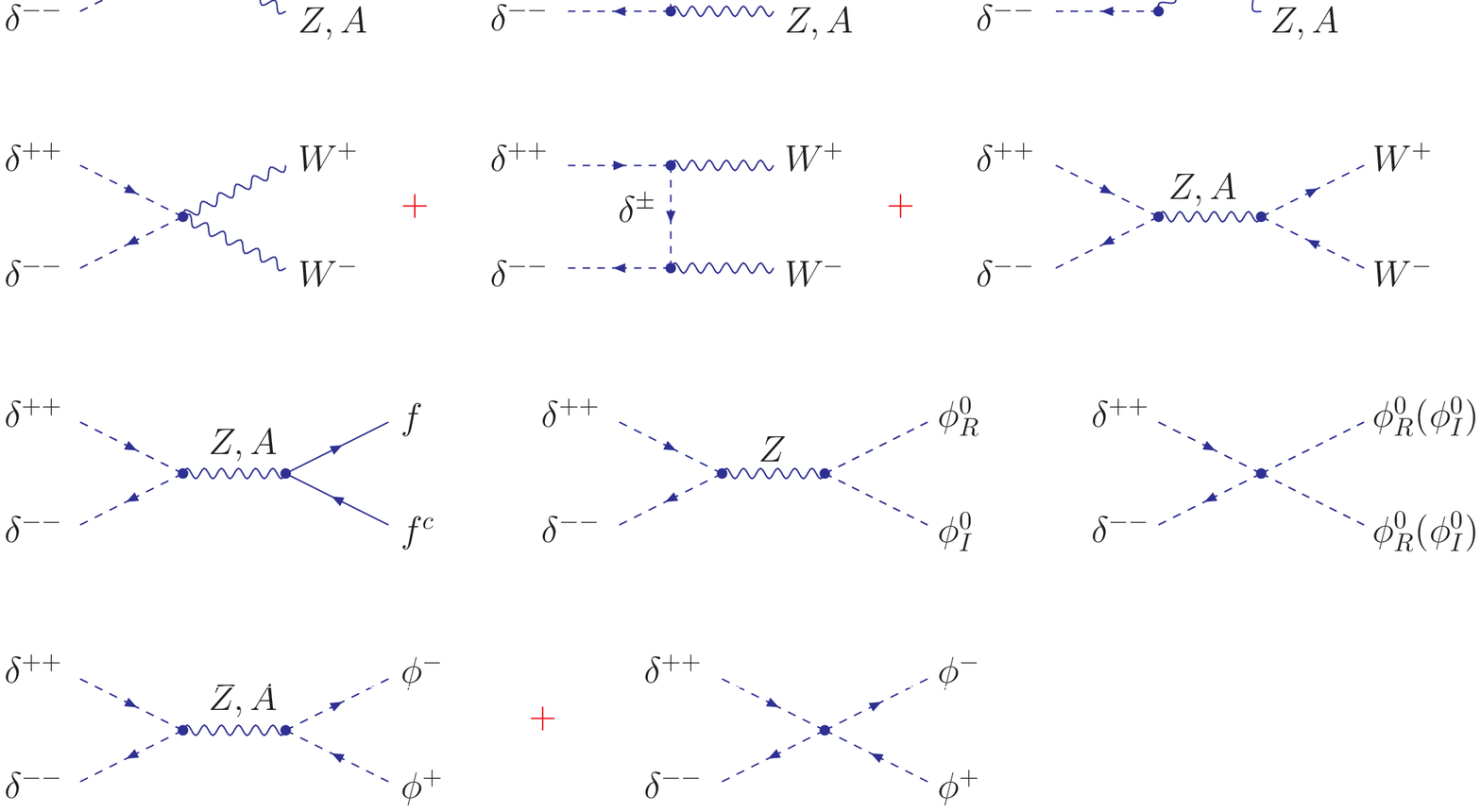}
  \vspace{-11.5cm} \caption{\label{d++d--} The $\delta^{\pm\pm}_{}\,+\,\delta^{\mp\mp}_{}$ annihilations.}
\end{figure*}

\clearpage
\newpage

\begin{figure*}[tbp]
\vspace{2cm}   \centering
  \includegraphics[width=15cm]{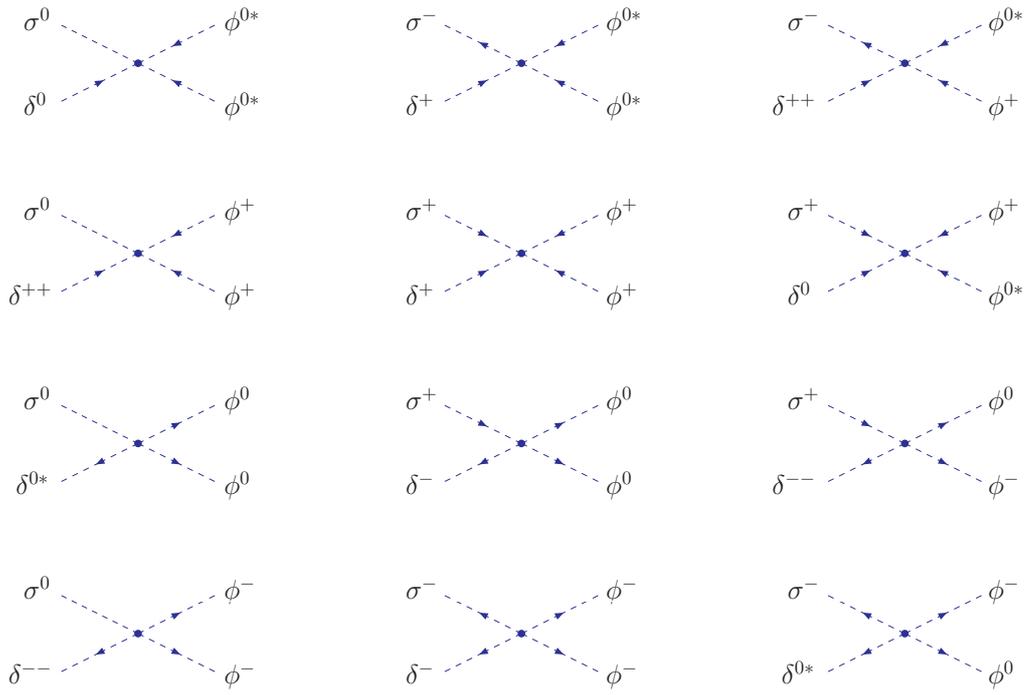}
  \vspace{-11.5cm} \caption{\label{sd} The $\Sigma\,+\,\Delta$ and $\Sigma\,+\,\Delta^\ast_{}$ annihilations.}\vspace{3cm}
  \end{figure*}

\clearpage
\newpage

\begin{figure*}[tbp]
\vspace{-5cm}   \centering
  \includegraphics[width=15cm]{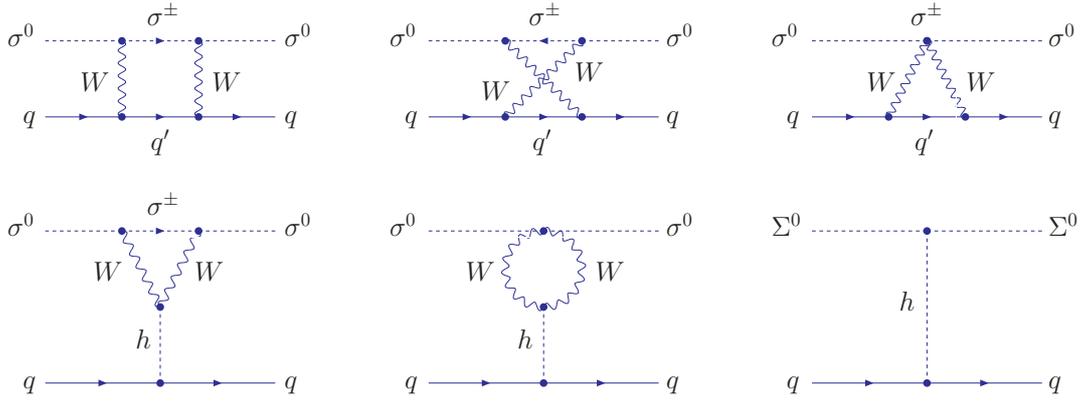}
  \vspace{-10.5cm} \caption{\label{s0n} The $\sigma^{0}_{} \,q \,\longrightarrow \,\sigma^0_{} \,q$ scattering.}
\end{figure*}

\begin{figure*}[tbp]
\vspace{-2cm}   \centering
  \includegraphics[width=15cm]{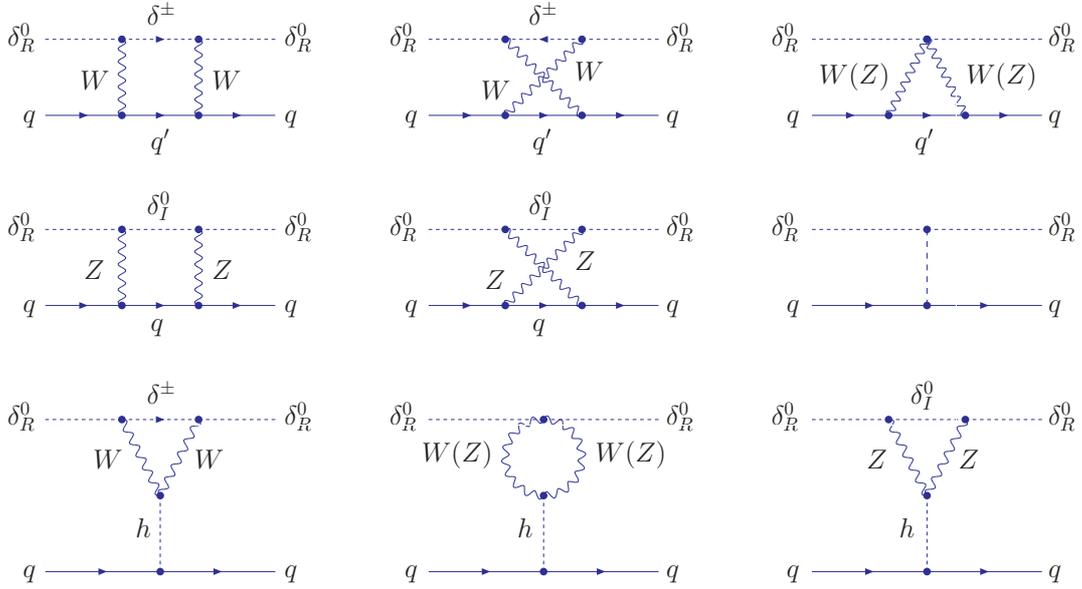}
  \vspace{-10.5cm} \caption{\label{d0n} The $\delta^{0}_{R}\,q \,\longrightarrow\, \delta^0_R\,q$ scattering.}
\end{figure*}

\begin{figure*}[tbp]
\vspace{-2cm}   \centering
  \includegraphics[width=15cm]{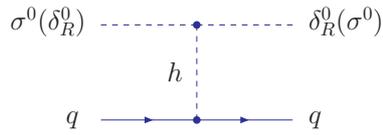}
  \vspace{-16.5cm}\caption{\label{s0d0} The $\sigma^0_{}\,q\,\longrightarrow\,\delta^0_R\,q$ and $\delta^0_R\,q\,\longrightarrow\,\sigma^0_{}\,q$ scattering.}
\end{figure*}

\end{document}